\documentclass[11pt]{article}

\usepackage{amsmath,amsfonts,amssymb,amsthm,bbm,stmaryrd}
\usepackage[dvips]{graphicx}
\usepackage{latexsym}
\usepackage{psfrag}
\usepackage[latin1]{inputenc}
\usepackage{hyperref}
\usepackage{color}
\usepackage{enumerate}
\numberwithin{equation}{section}
\usepackage{float}
 \usepackage{multirow}

\newcommand{\be}{\begin{equation}}
\newcommand{\ee}{\end{equation}}
\newcommand{\barray}{\begin{array}}
\newcommand{\earray}{\end{array}}
\newcommand{\bea}{\begin{eqnarray}}
\newcommand{\eea}{\end{eqnarray}}
\newcommand{\bs}{\begin{subequations}}
\newcommand{\es}{\end{subequations}}

\newcommand{\bit}{\begin{itemize}}
\newcommand{\eit}{\end{itemize}}
\newcommand{\bd}{\begin{description}}
\newcommand{\ed}{\end{description}}

\def\nn{\nonumber}

\def\la{\langle}
\def\ra{\rangle}

\def\w{\wedge}
\newcommand{\p}{\partial}
\newcommand{\na}{\nabla}

\newcommand{\R}{\mathbb{R}}

\newcommand{\f}{\frac}
\newcommand{\tl}{\tilde}

\renewcommand{\a}{\alpha} \renewcommand{\b}{\beta} \newcommand{\g}{\gamma}  
\renewcommand{\d}{\delta}  \newcommand{\eps}{\epsilon} 
 \renewcommand{\th}{\theta}  \newcommand{\vth}{\vartheta} 
    \renewcommand{\l}{\lambda}
\let\m=\mu    \let\n=\nu   \let\r=\rho \let\om=\omega
 \newcommand{\s}{\sigma}  \renewcommand{\t}{\tau}    
\let\G=\Gamma \let\D=\Delta    
 \let\Om=\Omega

\def\cL{{\cal L}}

\def\cN{{\cal N}}
\def\cB{{\cal B}}

\newcommand{\scri}{{\cal I}}

\newcommand{\ut}[1]{ \underset{\widetilde{}}{#1}{} }

\usepackage{slashed}
\newcommand{\sd}{\slashed{\delta}}

\newcommand{\eqons}{\,\hat{=}\,}
\newcommand{\eqon}[1]{\stackrel{{#1}}=}

\newcommand{\pbi}[1]{\underset{\leftarrow}{#1}}

\newcommand{\sscr}{\scriptscriptstyle\rm}

\newcommand{\Ger}{\r}

\begin{document}

\title{\bf Wald-Zoupas prescription with (soft) anomalies}

\author{\Large{Gloria Odak, Antoine Rignon-Bret and Simone Speziale}
\smallskip \\ 
\small{\it{Aix Marseille Univ., Univ. de Toulon, CNRS, CPT, UMR 7332, 13288 Marseille, France}} }
\date{\today}

\maketitle

\begin{abstract}
We show that the Wald-Zoupas prescription for gravitational charges is valid in the presence of anomalies and field-dependent diffeomorphism, but only if these are related to one another in a specific way. The geometric interpretation of the allowed anomalies is exposed looking at the example of BMS symmetries: They correspond to soft terms in the charges. We determine if the Wald-Zoupas prescription coincides with an improved Noether charge. The necessary condition is a certain differential equation, and when it is satisfied, the boundary Lagrangian of the resulting improved Noether charge contains in general a non-trivial corner term that can be identified a priori from a condition of anomaly-freeness. Our results explain why the Wald-Zoupas prescription works in spite of the anomalous behaviour of BMS transformations, and should be helpful to relate different branches of the literature on surface charges.
\end{abstract}

\tableofcontents

\section{Introduction}

The seminal Wald-Zoupas (WZ) paper \cite{Wald:1999wa} provides a prescription for the gravitational charges in the two cases of conservative boundary conditions and of leaky boundary conditions, with the latter that make the system non-conservative and thus the infinitesimal Hamiltonian generators non-integrable. It  reproduces the Arnowitt-Deser-Misner (ADM) charges for the Poincar\'e group at spatial infinity in the first case,  the 
Geroch and Dray-Streubel charges for the Bondi-van der Burg-Metzner-Sachs (BMS) group at future null infinity in the second case; It has  been recently extended to null hypersurfaces at finite distance and non-expanding horizons \cite{Chandrasekaran:2018aop,Ashtekar:2021kqj}. On the other hand, a more general framework has been developed by various authors in the last few years \cite{Harlow:2019yfa,Freidel:2020xyx,Chandrasekaran:2020wwn,Margalef-Bentabol:2020teu,Compere:2020lrt,Freidel:2021yqe,Freidel:2021cjp,Chandrasekaran:2021vyu,ciambelli2023asymptotic}.
In particular, \cite{Freidel:2021cjp,Chandrasekaran:2021vyu} have shown 
how to include in the covariant phase space arbitrary field-dependent diffeomorphisms and anomalies -- quantities whose field-space transformation under diffeomorphisms differs from the Lie derivative. 
Can the WZ prescription be applied in this more general context?
Answering this question is useful for future research, but also to better understand the precise relation between the recent literature and \cite{Wald:1999wa}.
In particular, both field-dependent diffeomorphisms and anomalies appear in the study of the BMS group;
How come Wald and Zoupas were able to derive the BMS charges without including either of these two features in their description?
Answering these two related questions motivates the analysis presented in this paper.

The answers lie in the fact that even if Wald and Zoupas didn't explicitly consider field-dependent diffeomorphisms and anomalies,  they made some precise assumptions about covariance. Our first result is to translate these assumptions to the formalism of  \cite{Freidel:2021cjp,Chandrasekaran:2021vyu}. We find that 
the WZ prescription can be applied also in the presence of a certain class of field-dependent diffeomorphisms and anomalies, contrarily to what one may initially expect,
but only provided these satisfy a precise relation relating one to the other.
We refer to the allowed anomalies as \emph{soft anomalies}. Their presence and the relation they satisfy are 
instrumental to understanding why the procedure works at future null infinity, thus answering our motivational question.

Our second result is to present a detailed comparison between the WZ prescription for the charges, and the prescription used in \cite{Harlow:2019yfa,Freidel:2020xyx,Chandrasekaran:2020wwn,Margalef-Bentabol:2020teu,Freidel:2021cjp,Chandrasekaran:2021vyu}, which we refer to as boundary-improved, or improved for short, Noether charges. These are always well-defined, so the question is whether they can be used to reproduce the WZ charges when the prescription for the latter is fulfilled. We find that a positive answer requires finding a corner Lagrangian satisfying a certain differential equation determined by the symmetry vector fields. 
When this equation can be solved, the WZ charges can be derived as improved Noether charges. Furthermore, the resulting Noether charges can be identified a priori, as those associated with a symplectic potential and boundary Lagrangian which are both anomaly-free: Anomalies can be present only in the corner terms.
What makes this construction possible is $(i)$ the existence of a covariant bulk Lagrangian given by the Einstein-Hilbert action or the tetrad action, and $(ii)$ the relation satisfied by the soft anomalies.
This result provides an independent definition of the WZ charges which fits very naturally in the language of \cite{Freidel:2021cjp,Chandrasekaran:2021vyu}. 

We complete our analysis looking at the four examples of spatial infinity, conservative boundary conditions on a time-like boundary, leaky boundary conditions on arbitrary null hypersurfaces and at future null infinity. In all cases we show how to reproduce the WZ expressions as 
improved Noether charges satisfying our criteria. The first three cases are rather trivial, but we include them because we believe they have useful pedagogical value. The non-trivial and most interesting example we consider is future null infinity. There we can see all the details of our analysis coming into play, understand how the allowed anomalies play a key role in order to get the standard BMS charges, and also explain how WZ were able to obtain them without addressing anomalies explicitly. This example further allows us to endow the WZ-allowed anomalies with a physical interpretation: They correspond to soft terms in the BMS flux-balance laws, those responsible for memory effects. One of the effects of the WZ prescription is to ensure that the anomaly contribution is entirely removed from the flux of the improved Noether charge, and placed instead in the definition of the charge. 
This example prompts us to refer to the anomalies allowed by the WZ prescription as soft. 

To be clear, the fact that the BMS charges can be derived as improved Noether charges is not new: it was already shown in \cite{Freidel:2021yqe} and \cite{Chandrasekaran:2021vyu}.  
These two papers however derive the required boundary Lagrangian and the corner shift \emph{a posteriori}, from the prior knowledge of the BMS charges. What we do here is instead deriving them \emph{a priori}, based on our criteria ensuing from a direct translation of the WZ conditions.

To conclude the introduction, let us briefly remind the reader of the general importance of including field-dependent diffeomorphism and anomalies in the covariant phase space, so to put potential future applications of our results in context. Field-dependent diffeomorphisms appear in various physical circumstances. As a first example, consider bulk extension of boundary symmetry vectors. It is often convenient to fix a specific extension, for instance Tamburino-Winicour \cite{Tamburino:1966zz} or Geroch-Winicour \cite{Geroch:1981ut}, or  preserving the gauge-fixing made by a bulk coordinate choice \cite{Barnich:2010eb,Compere:2018ylh,Henneaux:2018hdj}. These requirements make the extension field-dependent, and extensions can be relevant in the study of sub-leading charges, e.g. \cite{Godazgar:2018dvh,Hamada:2018vrw,Compere:2019gft,Freidel:2021ytz,Compere:2022zdz,Seraj:2022qyt}. Field-dependence of the boundary symmetry vectors themselves occurs in enlargements of the BMS symmetry \cite{Barnich:2010eb,Campiglia:2014yka,Compere:2018ylh,Freidel:2021yqe}, generically turning the Lie algebra into an algebroid \cite{Barnich:2010eb},
and in investigations of integrability using the `slicing' method, see e.g. \cite{Adami:2021nnf,Geiller:2021vpg}. 
Finally, field-dependent gauge parameters are familiar in the canonical approach, where they can be used to simplify the structure of the constraints, see e.g. \cite{Ashtekar:2020xll} for recent work using this idea. 
As for the anomalies \cite{Hopfmuller:2018fni,Harlow:2019yfa,Chandrasekaran:2020wwn}, they can appear whenever background structures 
 are present, and are central to the program of boundary observables in general. Taking them into account systematically helps us deepening our understanding of the covariant phase space and extending its applications. 
For instance, anomalies allow one to compute the cocyle of the Barnich-Troessaert bracket from first principles \cite{Freidel:2021qpz}, 
and explain the difference between the WZ charges and the Brown-York charges on null hypersurfaces \cite{Chandrasekaran:2021hxc}.
We hope that our paper will help shed light on how these various structures come together, and contribute to new applications of the formalism to address outstanding questions. In particular, in the study of extensions of the BMS symmetries and modifications of the charges that arise in that context \cite{Compere:2018ylh,Freidel:2021ytz,Compere:2022zdz,Campiglia:2020qvc}.

In the Appendix we give a brief review of the definition and evaluation of anomalies, and provide all explicit formulas relevant to the case of future null infinity. We also add some considerations on the subsequent derivations of the BMS charges that appeared in \cite{Barnich:2011mi,Flanagan:2015pxa,Grant:2021sxk}.

\bigskip

We use mostly-plus spacetime signature. Greek letters are for spacetime indices, and we will sometimes denote scalar products by a dot.
When needed, lower case latin letters $a,b,...$ are hypersurface indices, and upper case latin letters $A,B,...$ are indices for the 2d cross-sections of the hypersurface.
In all cases, $(,)$ denotes symmetrization, $\la,\ra$ trace-free symmetrization, and $[,]$ antisymmetrization. An arrow $\leftarrow$ under a $p$-form or $p$-form index means pull-back, and $\eqons$ means on-shell of the field equations. 
We use units $16\pi G=c=1$.

\section{Charge prescriptions}

The starting point for the covariant phase space is a symplectic potential $\th$ related to the variation of the Lagrangian by
\be\label{dL}
\d L \eqons d\th,
\ee
from which one reads the (pre)-symplectic 2-form current $\om:=\d\th$, which satisfies $d\om\eqons 0$. We use small greek letters for the currents, namely the integrands, and capital letters for the integrated quantities. However, we will loosely speak of both as symplectic potentials and 2-form, for ease of language.
We are interested in the general situation that  
includes field-dependent diffeomorphisms, hence $\d\xi\neq 0$, and anomalies, namely non-covariant quantities 
whose field space transformation under a diffeomorphism differs from the Lie derivative. 
We follow \cite{Freidel:2021cjp} for notation and the general framework, and define the anomaly operator 
\be\label{Dxidef}
\D_\xi:= \d_\xi-\pounds_\xi-I_{\d\xi}.
\ee
Here $\pounds_\xi$ is the Lie derivative in spacetime, and $\d_\xi$ the Lie derivative in field space.
See Appendix~\eqref{AppA} for more details and the definition of the field-space inner product $I_{\d\xi}$. This term only acts on field-space forms.
We call \emph{covariant} a field-space quantity such that $\d_\xi=\pounds_\xi$. We see from \eqref{Dxidef} that this amounts to a vanishing anomaly for field-space scalars. But for field-space forms, covariance in the presence of field-dependent diffeomorphisms may require a non-vanishing anomaly, compensating the action of $I_{\d\xi}$. Accordingly, it is $\D_\xi+I_{\d\xi}$ that measure the non-covariance, and not the anomaly operator alone.
The lack of covariance can occur in the presence of background structures which are described by spacetime fields but constant under variations in field space, and it will be necessary to understand the charges on null hypersurfaces. 
The only restriction made in \cite{Freidel:2021cjp} and also here, is that the non-covariance of the Lagrangian should at most be a boundary term, namely that there are choices of Lagrangians such that the background structure that may lead to a breaking of covariance only enters through boundary terms.\footnote{This includes the treatment of anomalous bulk Lagrangians like ADM, since it differs from the covariant Einstein-Hilbert Lagrangian by a boundary term.}
Accordingly, $L=L^{\sscr cov}+d\ell$ with $\D_\xi L^{\sscr cov}=0$ and $\D_\xi L=d a_\xi$, and  we define the Lagrangian and symplectic anomalies $a_\xi$ and $A_\xi$ via
\be\label{anomalies}
\Delta_\xi \ell = a_\xi, \qquad \Delta_\xi \theta = \delta a_\xi -a_{\d \xi} + d A_\xi.
\ee
If such anomalies are present, they show up in the formula for the Noether charges as well as for the Hamiltonian generators: Following the standard procedure \cite{Iyer:1994ys} but allowing for non-vanishing anomalies and field-dependent diffeomorphisms, one obtains \cite{Freidel:2021cjp}
\begin{align}\label{jN}
& j_\xi:= I_\xi\th-i_\xi L-a_\xi \eqons d q_\xi, \\
\label{Ixiom}
&\sd d h_\xi:= -I_\xi \om \eqons d\left(\d q_\xi -i_\xi\th-q_{\d\xi}-A_\xi\right).
\end{align}
The Lagrangian and symplectic anomalies enter respectively the Noether charge $q_\xi$ and (infinitesimal) Hamiltonian generator $dh_\xi$.
Notice that the Hamiltonian generator depends only on $\th$, whereas the Noether charge depends on $\th$ but also explicitly on the boundary Lagrangian via its anomaly.

To understand the meaning of $a_\xi$ in the first formula, consider the pull-back on a given hypersurface. 
If the hypersurface is a boundary used to define the covariant phase space, then the relevant symmetry vectors $\xi$ are those tangent to it, since they are the only ones preserving the boundary and thus the phase space. Then the pull-back of $i_\xi L=0$ vanishes, and the variation of the Noether charge along the boundary has two contributions: one is the symplectic flux of the symmetry, and this is the contribution due to physical degrees of freedom crossing the hypersurface. The other is the anomaly. This term induces a charge variation caused by the background structure, thus introducing a non-dynamical contribution to the flux. For instance, this term is non-zero if one uses a normal that depends on the foliation to which the boundary belongs. \footnote{A different situation occurs if the background structure breaks diffeomorphism invariance entirely, for instance if we have matter fields but the (curved) metric is treated as a fixed background, the anomaly term $a_\xi$ is nothing but the energy-momentum tensor of matter, and one recovers the non-general-covariant notion of bulk Noether charge. This observation allows one to reverse the standard viewpoint that sees Noether charges as global, becoming surface charges in the special case of local gauge symmetries; and consider instead that all Noether charges are surface charges, becoming global only in the presence of anomalies introduced by background structures \cite{Antoine}}.

In the second formula, $A_\xi$ contributes as an additional potential obstruction to integrabillity.
If the right-hand side of \eqref{Ixiom} is not integrable, the diffeomorphism transformation fails to be a Hamiltonian vector field, whence the thermodynamical notation $\sd$ \cite{Barnich:2007bf}. Non-integrability happens for instance in the presence of a lateral boundary $\cB$ joining two space-like slices, because in this case the property $d\om\eqons 0$ is not enough to guarantee that the symplectic form is conserved between the two space-like slices.  When there is flux leaking through the lateral boundary, \eqref{Ixiom} will in general not define a Hamiltonian generator, and one needs a prescription for the charges. The obvious choice of taking the Noether charge via \eqref{jN} leads for the Einstein-Hilbert action to the Komar formulas. These have various useful properties, but also shortcomings that have been known for a long time (such as wrong factors of 2 in the energy at both spatial and future null infinity, generically non-invariant flux-balance laws,
and so forth, see e.g. \cite{Iyer:1994ys,Szabados:2009eka}), hence the motivation for a different prescription.\footnote{
The literature contains various interesting proposals on how to achieve integrability, for instance enlarging the phase space introducing embedding fields that move the boundary in such a way that the outgoing flux is absorbed into the definition of the charge  \cite{Ciambelli:2021nmv,Freidel:2021dxw}. Another approach uses a new Leibnizian bracket with respect to which the charges are integrable \cite{Kabel:2022efn}. At least in non-dynamical cases, there exists also the possibility of obtaining integrability finding an appropriate field-dependence of $\xi$ so that the $q_{\d\xi}$ term cancels the obstruction, a procedure known as `slicing', see e.g. \cite{Adami:2021nnf}. We will not consider these alternative constructions here, and only discuss the WZ prescription.}

In order to understand the prescriptions, it is important to recall that the covariant phase space constructed above is not unique, because of the existing freedom in choosing the symplectic potential. This freedom is two-fold  \cite{Jacobson:1993vj}: first, given any $\theta$ satisfying \eqref{dL} we can add to it any spacetime exact 3-form $d\a$; second, we can add a boundary term to the Lagrangian, which doesn't change the field equations nor the symplectic structure. 
These two cohomological ambiguities (one in spacetime and one in field space) are summarized by 
\begin{align}\label{freedom}
& L\rightarrow L+dY, \qquad \th\rightarrow \th+d\a+\d Y, \qquad \om\rightarrow\om+\d d\a.
\end{align}
In particular, the Noether charge defined via \eqref{jN} depends on the choice of representative, and transforms as
\be
q_\xi \rightarrow q_\xi + i_\xi Y + I_\xi\a. \label{qfreedom}
\ee

For a given $L$, we refer to the choice of $\th$ obtained simply removing $d$ as the `bare' choice. This choice follows if the symplectic potential  is defined using Anderson's homotopy operator \cite{Iyer:1994ys,Anderson:1996sc,Barnich:2000zw,Barnich:2001jy,Freidel:2020xyx}, which is the approach taken in \cite{Freidel:2021cjp}. Another mathematical way to eliminate the freedom is to require the Noether current \eqref{jN} to be weakly vanishing \cite{Barnich:2001jy}. 
These choices are convenient for bookkeeping and can always be made, but they are however \emph{not needed} to obtain the results used and derived here. In the rest of the paper, we will consider arbitrary $\th$'s, without any a priori mathematical prescription.

\subsection{Improved Noether charge from phase space polarization}\label{SecImproved}

The main idea that we would like to recall from the WZ paper is that one should resolve the ambiguities in the definition of the charges by deciding under which physical requirements they are to be conserved.
Mathematically, this can be controlled trading the initial symplectic potential $\th$ (be it the bare one or any other chosen one) for a symplectic potential such that its pull-back on the lateral boundary $\cB$ vanishes in the subset of the phase space corresponding to a desirable physical requirement, such as a choice of conservative boundary conditions, or a choice of stationarity conditions.  
In practise,
 one takes the pull-back on the lateral boundary and decomposes it as follows, 
\be\label{th'}
\pbi{\th} = \th' - \d\ell + d\vth,
\ee
where $\th'$ is required to be in the form $p\d q$ for some choice of polarization of the phase space.
The new $\th'$ corresponds to $L':=L+d\ell$, namely a theory with the same field equations, and is equivalent to $\th$ under the freedom \eqref{freedom}.
The idea of changing from the initial $\th$ to a physically motivated $\th'$ dates back to \cite{Iyer:1995kg} and \cite{Wald:1999wa}, was generalized in \cite{Compere:2008us} and \cite{Harlow:2019yfa} to include the corner potential $\vth$, and takes a central role in various follow-up works \cite{Freidel:2020xyx,Chandrasekaran:2020wwn,Margalef-Bentabol:2020teu,Freidel:2021cjp,Chandrasekaran:2021vyu}.

The terms $\ell$ and $\vth$ appearing above are produced by the manipulations needed to put (the pull-back of) ${\th}$ in the chosen $\th'$ form. 
The explicit form of $\vth$ depends also on the representatives chosen for $\th$ and $\th'$.
Since we require $\th'=p\d q$, $\ell$ is manifestly the boundary term to be added to the Lagrangian to have a well-defined variational principle with those boundary conditions. However \eqref{th'} does not identify a unique $\ell$, since the condition is still satisfied under the replacement
\be\label{corneramb} 
(\ell,\vth)\rightarrow(\ell+dc,\vth+\d c). 
\ee
Therefore, for a given representative $\th$, one can compute a unique $\vth$ only once a choice for $\th'$ \emph{and} $\ell$ is made.\footnote{
Unique $\vth$ up to addition of exact 2-forms, but these will be irrelevant in the following since we will only look at compact corners. 
Accordingly, we will ignore all $2d$-exact forms in the rest of the paper.
Notice also that fixing both $\th'$ and $\ell$ can be equivalently seen as fixing $\th'$ and $\vth$. This is the viewpoint taken in \cite{Chandrasekaran:2021vyu}, where the chosen quantities are referred to respectively as boundary and corner (symplectic) fluxes. The freely choosable $\vth$'s have to be related by \eqref{corneramb}, just like the freely choosable $\th$'s have to be related by \eqref{th'}.}
For instance, a non-vanishing $\vth$ occurs for Dirichlet boundary conditions if $L$ is the Einstein-Hilbert Lagrangian and $\ell$ is the Gibbons-Hakwing-York term,  as established as early as \cite{Burnett:1990tww}. As there observed, the resulting $\vth$ shifts the symplectic 2-form,
\be\label{omprime}
\om'=\d\th'=\om-d\d \vth.
\ee

We want to characterize the physical situations in which the new symplectic potential vanishes on the lateral boundary $\cB$, namely
\be\label{th'0}
\th'\eqon{\sscr\cB} 0.
\ee
Since it is in the form $\th'=p\d q$, we can distinguish two cases, depending on whether it is $\d q$ or $p$ to vanish, and which we name following \cite{Wald:1999wa}.

\bit
\item[Case I:] We impose \emph{conservative} boundary conditions $\d q\eqon{\sscr\cB}0$. 
The new symplectic 2-form $\om'$
also vanishes on the lateral boundary, 
\be
\pbi{\om}\eqon{\sscr\cB}0,
\ee
and therefore is preserved between the initial and final space-like hypersurfaces. This makes the system conservative, hence the name.
In other words, the system is in case I within each cotangent space at fixed $q$, but not for trajectories that vary both $p$ and $q$.

Clearly, different choices of conservative boundary conditions are possible, 
corresponding to different choices of polarizations,
and this turns out to affect the charges. 
Most literature focuses on Dirichlet boundary conditions, but the charges obtained from Neumann and York boundary conditions were computed in \cite{Odak:2021axr} for a time-like lateral boundary. They turn out to be different, and in agreement with what can be computed using $3+1$ canonical methods. See \cite{nBYus} for an exploration of alternative boundary conditions in the case of a null boundary.

\item[Case II:] There exist solutions for which $p\eqon{\sscr\cB}0$. They provide a notion of \emph{stationary} backgrounds, whose precise nature depends on the form of $p$, namely on the polarization chosen. We can distinguish two situations, one in which all $p$'s vanish, and one in which only some vanish, and \eqref{th'0} is achieved by the vanishing of the complementary $\d q$'s. Either way, the symplectic 2-form $\om'$ is not conserved, 
\be\label{omrad}
\pbi{\om}\stackrel{\sscr\cB}{\neq}0,
\ee
because there are no restrictions on the variations $\d p$ and at least some of the  $\d q$'s. Therefore these are radiative or \emph{leaky} boundary conditions.

\eit

It should be stressed that we are making these characterizations with the goal of resolving the ambiguities in the charges, and not of restricting the phase space. Once the corresponding $\th'$ is chosen, we compute the associated charges, and then we use them in the \emph{full} phase space. This means that charges defined using conservative boundary conditions will not be conserved in the full phase space, and charges defining using a specific stationarity condition will not be conserved when evaluated around any other solution not respecting it. Clearly, charges constructed using the different perspectives of Case I and II but corresponding to the \emph{same} polarization are equal and have equal properties.

The formulas for charges associated with the new symplectic potential $\th'$ take exactly the same form as before, namely \cite{Freidel:2021cjp}
\begin{align}\label{iNflux}
& j'_\xi := I_\xi\th'-i_\xi L'-a'_\xi \eqons dq_\xi', \\ \label{Iom'}
& \sd d h_\xi:= -I_\xi \om' \eqons d\left(\d q'_\xi -i_\xi\th'-q'_{\d\xi}-A'_\xi\right),
\end{align}
where $L'=L+d\ell$ and $\om'=\d\th'$. In other words, the formalism allows one to treat all choices on equal footing. 
Again, the infinitesimal Hamiltonian generator depends only on $\th'$, whereas the Noether charge depends on $\th'$ but also explicitly on the boundary Lagrangian $\ell$ via its anomaly: $(\th',\ell)\mapsto q'_\xi$. In particular, $q'_\xi$ depends on any corner term that may be present in the choice of $\ell$, which is not visible from $L'$ and $\th'$. 
The relation between the Noether charges associated to $(\th',\ell)$ and the initial ones is\footnote{The reader may notice a notational hiccup at this point. Logically, it would make more sense to denote the boundary Lagrangian $\ell'$, so that one can use $\ell$ to refer to whatever choice of corner was present in the initial $q_\xi$. Accordingly, one should add primes on both $\ell$ and $\vth$ on the right-hand side of \eqref{th'}, and following formulas. We choose not to do so and instead follow the notation of \cite{Freidel:2021cjp}. This allows us to keep the notation lighter, and also refer to that paper for all proofs. Notice also that the practical use we will make of the unprimed notation will be to specialize to the bulk covariant Lagrangian with no boundary term, so no confusion will arise as to what $\ell$ refers to. }
\be\label{iN}
q_\xi'=q_\xi+i_\xi\ell-I_\xi\vth.
\ee
Keeping the primed notation is useful if we want to compare boundary-improved charges to specific bare charges.
For instance, take $L$ to be the Einstein-Hilbert Lagrangian, and $\th$ its bare symplectic potential. Then $q_\xi$ is the original Noether charge \cite{Iyer:1994ys}, given by the Komar formulas and their limitations. If we add the boundary Lagrangian $\ell$ given by the Gibbons-Hawking-York term and choose the Dirichlet polarization for $\th'$, the improved Noether charges give the Brown-York formulas \cite{Iyer:1995kg,Harlow:2019yfa}.
In \cite{Odak:2021axr} we referred to this prescription as Freidel-Geiller-Pranzetti formula, since we used the notation of \cite{Freidel:2020xyx}, but given the number of authors contributing to these developments, it seems fair to simply talk about improved Noether charges. 
The improvement with respect to the original, `bare' Noether charges, is two-fold. First, the Brown-York formulas give the correct ADM charges at spatial infinity \cite{Brown:1992br}, unlike the Komar formulas. 
Second, \eqref{iN} can be made invariant under the cohomological ambiguities \eqref{freedom}
\cite{Margalef-Bentabol:2020teu,Chandrasekaran:2021vyu}.
Indeed, if we require that the choice of polarization $\th'$ is {kept fixed} under \eqref{freedom}, we have $(\ell,\vth)\rightarrow(\ell-Y,\vth+\a)$, therefore even if $q_\xi$  changes as in \eqref{qfreedom}, $q'_\xi$ is invariant. In other words, it is the prescription of working with a unique $\th'$ that eliminates these ambiguities. 

On the other hand, fixing $\th'$ alone is not sufficient to obtain a unique charge, because as anticipated above, $q_\xi'$ depends also on the boundary Lagrangian chosen. This can be seen explicitly observing that \eqref{iN} is affected by the corner ambiguity \eqref{corneramb}, which leads to \cite{Chandrasekaran:2021vyu}
\be\label{qc}
q'_\xi\rightarrow q_\xi'-\D_\xi c.
\ee
Therefore even if the cohomology ambiguities $(Y,\a)$ are fixed by the choice of $\th'$, there is still an ambiguity in the charge if anomalies are present. 
This ambiguity is removed if one does not prescribe only $\th'$ but also a specific choice of $\ell$, thus fixing $c$.\footnote{This may be taken as a suggestion that what matters to get unique charges is prescribing a specific action principle 
including boundary and corner terms, as pointed out in \cite{Margalef-Bentabol:2020teu,Chandrasekaran:2021vyu}.
However more work is needed in our opinion before this suggestion is borne out, because counter-examples exist, both ways. Going one way, one can think of the example of adding an anomalous corner term (hence changing the charge) but which is globally defined (hence not entering as corner terms in the action principle). Going the other way, the example of time-like boundaries with non-orthogonal corners reviewed below in Section~\ref{SecTimelike} shows that there is no corner shift needed to get the BY charges corresponding to the WZ prescription even though there are corner terms in the action principle. At least in the case of the WZ prescription, what we will find is that the corner shift is not related to corner terms in the action principle, but rather in removing anomalies from the boundary Lagrangian.}
 
Let us now discuss how the restriction \eqref{th'0} affects integrability, and the ensuing relation between the improved Noether charge and the Hamiltonian generator.
Consider first the familiar case of no anomalies and $\d\xi=0$.
When \eqref{th'0} holds, the improved Noether charge $q_\xi'$ is conserved (this follows because the $\xi$'s allowed in the covariant phase space are tangent to the boundary, and thus the second term in \eqref{iNflux} vanishes taking the pull-back). 
The Hamiltonian generator \eqref{Iom'} is integrable, and the Hamiltonian coincides with $q'_\xi$, up to constant terms in field space. Such terms can be fixed for instance looking at a reference solution \cite{Wald:1999wa}, requiring the Hamiltonian charges to vanish there. 
Having established this, one can take the prescription of using the improved Noether charges in the full phase space. This prescription gives charges that by definition have the useful property of being conserved and Hamiltonian generators in the conservative or stationary subspace.
Notice that this prescription is equivalent to defining the charges starting from the Hamiltonian generator and subtracting the flux, since $-I_\xi \om' + di_\xi\th' = d\d q'_\xi $. The important point to stress is that $q'_\xi$ is not associated to an arbitrary choice of symplectic potential, but to the physically preferred $\th'$. This removes any ambiguity in the procedure.

In the general case with $\d\xi\neq 0$ and anomalies, we can compare the Hamiltonian generator and the improved Noether charge as follows.
When \eqref{th'0} holds, the Hamiltonian generators are integrable iff there exists a functional $X$ such that
\be\label{intcond}
d Y=\d X,
\ee 
where
\be
Y=-q'_{\d\xi}-A'_\xi.
\ee
This requirement means that 
\be\label{Xds}
X=ds_\xi+C_\xi, \qquad \d s_\xi = Y,
\ee
and $C_\xi$ is a constant in field-space. If this condition is satisfied, we can again prescribe the charges on the full phase space subtracting the symplectic flux, via
\be\label{genWZ}
\d d h_\xi {:=} -I_\xi \om' + di_\xi\th' \, \hat{=}\, 
d\left(\d q'_\xi -q'_{\d\xi}-A'_\xi\right) = \d d (q'_\xi+s_\xi).
\ee
The first equality follows from \eqref{Iom'}, and the second from \eqref{Xds}. This formula provides a definition for the Hamiltonian charge associated to the physically selected $\th'$, and works only if the anomalies satisfy the descent-type equation \eqref{intcond}. This is not yet the WZ prescription but a generalization thereof, since as we will review in the next Section, the WZ prescription makes additional requirements than just a specific $p\d q$ form of $\th'$. 

From the definition \eqref{genWZ} it follows that
\be\label{hqs}
h_\xi = q'_\xi +s_\xi,
\ee
up to field-space constants as before (the constant $C_\xi$ above drops out on the other hand).
Because of the extra term $s_\xi$, the prescription \eqref{genWZ} associated with the chosen $\th'$ does \emph{not} coincide in general with the improved Noether charge $q'_\xi$ associated with a given $(\th',\ell)$. 
However, there are two interesting remarks to make at this point. 
First, the formula \eqref{Iom'} is invariant under the corner Lagrangian shift \eqref{corneramb}, unlike the improved Noether charge which changes according to \eqref{qc}. Therefore, we can change the boundary Lagrangian by a corner term without affecting the Hamiltonian generator, and use this freedom to find a corner-improved Noether charge that matches the Hamiltonian charge. 
In other words, one can ask whether there is a choice of $\ell$ compatible with \eqref{th'} such that its Noether charges match the Hamiltonian prescription.
Comparing \eqref{qc} with \eqref{hqs}, we see that the matching is possible if there exists a corner term $c$ whose anomaly reproduces the integrable anomalies appearing in \eqref{Xds}, namely
\be\label{sDc}
\D_\xi c =-s_\xi.
\ee
If such $c$ exists, the corner-improved Noether charge associated with $\th^{\sscr c}=\th'$ and $\ell^{\sscr c}=\ell+dc$ matches the Hamiltonian charge,\footnote{The notation $^{\sscr c}$ stands for corner-improved, and should not be confused with the notation for covariant used in \cite{Chandrasekaran:2021vyu}. We don't use any specific notation for covariant quantities, although typically we will associate them with the initial, unprimed quantities.}
\be
q^{\sscr c}_\xi = q'_\xi+s_\xi\equiv h_\xi.
\ee
To be precise, the last equivalence is only up to the field-space constants mentioned above, since these can be freely added to $h_\xi$ in order to satisfy special vanishing requirements, but not to $q^{\sscr c}_\xi$ which is defined uniquely. 
The condition \eqref{sDc} is a partial differential equation that relates the corner improvement to the allowed anomalies of the symmetry vectors $\xi$. 
We will see below in the case of future null infinity an example of this equation and of its solution. In general, we don't know whether it is always possible or not to solve it. Whenever it is, the generalized WZ charge \eqref{genWZ} can always be derived as an improved Noether charge.
We will show in the next Section that the WZ additional requirements allows us to get a more explicit form for $s_\xi$, and we will make more comments about solving it then. 

The second remark is that the flux of this corner-improved Noether charge is still anomalous, since it is given by
\be
\pbi{d q}{}^{\sscr c}_\xi = I_\xi \th^{\sscr c} -a^{\sscr c}_\xi= I_\xi \th' -a'_\xi +ds_\xi.
\ee
This provides also the flux of the Hamiltonian charge (up to the usual field-space constants), since the $\d$-variation of the above expression must match \eqref{genWZ}. 
Therefore, the charges are \emph{not} automatically conserved when \eqref{th'0} holds.
Clearly, additional physical requirements could be useful to achieve conservation when $\th'$ vanishes on the lateral boundary. We will see next that the WZ prescription provides precisely such missing requirements, by forcing $a^{\sscr c}_\xi=0$. 
As a result, $I_\xi\th'$ gives the flux responsible for the variation of the charges, and one obtains charges that are conserved under the desired circumstances for which $\th'$ vanishes.\footnote{Even if the variation of the charges is given in the end by $I_\xi\th'$, it is still preferable to characterize the physical requirements such as stationarity in terms of $\th'$, as requirements on $I_\xi\th'$ may be ambiguous. We will see in \cite{nBYus} an example of such ambiguity.}

\subsection{Wald-Zoupas prescription}\label{SecWZ}

We now review the WZ prescription from \cite{Wald:1999wa} and highlight the additional inputs that are brought in with respect to the previous general discussion. 
The prescription is based on removing the radiative part from the symplectic flux, identified making use of a background structure that can be attributed to the lateral boundary. 
To do so, one selects a symplectic potential $\bar\th$ based on three criteria:
\bit
\item[0.] It must be a potential for the pull-back of the symplectic 2-form on the boundary, namely
\be\label{WZpb}
\underset{\leftarrow}\om=\d\bar\th.
\ee
\item[1.] It must be a local and covariant functional of the dynamical fields and background structure.
This is sometimes assumed to imply vanishing anomalies and field-independent diffeomorphisms, but we will see shortly that it is more general than that -- and this is crucial to understand the future null infinity results.
\item[2.] It should vanish for conservative boundary conditions, case I presented earlier, or for arbitrary perturbations around stationary solutions, case II. The latter means that it must be of the form $F(g)\d g$ where $F(g_{\rm stationary})=0$. In reference to the earlier discussion, if we think of $\bar\th$ as a certain $p\d q$ polarization, then WZ stationarity is of the type $p=0$.

\eit

Ideally, these criteria should be enough to single out a unique choice for $\bar\th$, and this is indeed the case in the examples 
that we will review below.
In case II, the preferred $\bar\th$ satisfying all criteria is identified as the radiative symplectic flux, 
namely, a quantity whose vanishing means that all metrics sharing the background structure agree that the solution is stationary.
As a consequence of the requirements made, one typically obtains
\be\label{thbar}
\bar \th = \pbi{\th} + \d b,
\ee
for some non-vanishing $b$ defined on $\cN$. 
We can interpret this formula as a special case of \eqref{th'}, where $\th'=\bar\th$ has to satisfy the WZ requirements above, $b$ is the pull-back of a boundary Lagrangian up to the corner ambiguity \eqref{corneramb}, and $\vth$ vanishes or is at most a total variation so that it can be reabsorbed in $b$. An arbitrary $d\vth$ cannot be present because it would violate 
\eqref{WZpb} hence condition $0$.

The WZ prescription for the integrable charges is then to subtract the radiative flux on $\cN$,
\begin{align}\label{WZdef}
\sd \pbi{{{d} q}}{}^{\sscr WZ}_\xi {:=} 
-I_\xi \om + di_\xi \bar \th
\,\hat =\, d\left(\d\bar q_\xi-\bar q_{\d\xi}-\bar A_\xi\right).
\end{align}
This is the same formula that we discussed in the previous Section. The novelty is the additional restriction given by conditions 0 and 1.
To study the most general situation under which the WZ prescription works (namely if we can replace $\sd$ with $\d$),
let us look closely at the covariance requirement.
 This property, as spelled out in footnote 9 of \cite{Wald:1999wa}, translates in our notation to\footnote{
Referring to the background fields as $\chi$ and the dynamical fields as $\phi$, the requirement spelled out in that footnote is $\bar\th(\chi,\varphi_*\phi,\varphi_*\d\phi)=\varphi_*\bar\th(\chi,\phi,\d\phi)$, and assumes the transformation law $\d\phi\mapsto \varphi_*\d\phi$. 
This transformation law is fine if the diffeomorphism is field-independent, but if it is field-dependent, one has to use $\d\phi \mapsto \d (\varphi_*\phi)$ in order for
the total system background+perturbation to be physically equivalent after the diffeomorphism. Accordingly, the  WZ requirement should be modified to
$\bar\th(\chi,\varphi_*\phi,\d (\varphi_*\phi))=\varphi_*\bar\th(\chi,\phi,\d\phi)$.
At the linearized level, this gives 
\[
\bar\th(\chi,\pounds_\xi\phi,\d\phi)+\bar\th(\chi,\phi,\d\pounds_\xi \phi) = \pounds_\xi\bar\th(\chi,\phi,\d\phi)=\bar\th(\pounds_\xi\chi,\phi,\d\phi)+\bar\th(\chi,\pounds_\xi\phi,\d\phi)+\bar\th(\chi,\phi,\pounds_\xi\d\phi),
\]
from which \eqref{WZfootnote} follows. In a previous version of the paper on arXiv, we considered a stronger condition in which $\D_\xi\bar\th$ and $I_{\d\xi}\bar\th$ vanish  individually. This is not necessary, and the correct version leads us to simpler equations when applied to the BMS analysis.
 }
\be\label{WZfootnote}
 \D_\xi\bar\th + I_{\d\xi}\bar\th=0 \qquad \Leftrightarrow\qquad (\d_\xi - \pounds_\xi) \bar\th = 0.
\ee
In other words, the anomaly of $\bar\th$ and the allowed field-dependent diffeomorphisms are constrained, so that $\bar\th$ is covariant: its field-space derivative coincides with the spacetime Lie derivative.
 This condition is indeed sufficient to guarantee integrability, since we can rewrite \eqref{WZdef} using
\be
 -I_\xi \om + di_\xi \bar \th = \d I_\xi\bar \th -\D_\xi\bar\th -I_{\d\xi}\bar\th \qquad \Rightarrow \qquad
\sd \pbi{{{d} q}}{}^{\sscr WZ}_\xi {=}\, \pbi{\d dq}{}^{\sscr WZ}_\xi {=}\,  \d I_\xi\bar\th, 
\ee 
namely
\be\label{WZflux}
\pbi{{d} q}{}^{\sscr WZ}_\xi {=}  I_\xi\bar\th
\ee
up to (spacetime exact) field-space constants, that can be used as described earlier if one needs to set the charges to zero for a specific reference solution.

On the other hand, the WZ covariance requirement does \emph{not} imply that all anomalies vanish. 
However, it implies some restrictions. Indeed, we know that :
\begin{subequations}\label{mild0}\begin{align}
& \D_\xi\bar\th= \D_\xi\pbi{\th}+\d\D_\xi b-\D_{\d\xi}b=\d \bar a_\xi -\bar a_{\d\xi} +d \bar A_\xi  \\
& \pbi{I_{\d\xi}\bar\th} =  d\bar q_{\d\xi} + a_{\d\xi}.
\end{align}\end{subequations}
In the first we used $\underset{\longleftarrow}{\D_\xi\th}=\D_\xi{}\pbi{\th}$ which is valid for tangent $\xi$.
Hence, \eqref{WZfootnote} gives :
\be
\D_\xi\bar\th +  I_{\d\xi}\bar\th = \d \bar a_\xi + d (\bar A_\xi + \bar q_{\d\xi}) = 0
\ee
 or :
\be\label{zidane}
\d \bar a_\xi = -d(\bar q_{\d\xi}+\bar A_\xi).
\ee
In the above formula one can freely replace $\bar A_\xi$ with $ A_\xi$, since anyways there is no corner difference between $\bar \th$ and $\th$.
This relation is a special case of \eqref{intcond}, in which $X$ is determined by the Lagrangian anomaly $\bar a_\xi$. 
As a consequence, 
\be
\bar a_\xi =d s_\xi +C_\xi, \qquad \d s_\xi=-\bar q_{\d\xi}-A_\xi.
\ee

We see that the WZ covariance requirements \eqref{WZfootnote} are compatible with the presence of field-dependent diffeomorphisms and anomalies, provided $\d\xi$, $a_\xi$ and $\D_\xi b$ are related by \eqref{zidane}. 
We refer to such WZ-compatible anomalies as {mild} or \emph{soft} anomalies. We will see below an example that justifies this name. 
This is the most general situation allowed by the WZ requirements, and as seen above it is enough to guarantee integrability of their prescription for the charges. If we compare with the generalized WZ prescription \eqref{genWZ}, we see that \eqref{zidane} is a special case of the integrability condition \eqref{intcond}. The restriction comes from having added conditions 0 and 1.

We stress that we have done nothing new concerning  the charges: we have merely re-derived the same formula of WZ, namely \eqref{WZflux}, under the same conditions as they did. Our only contribution is to point out that such conditions, and therefore the derivation, do admit anomalies, provided they are soft in the above sense.

Now that we have clarified that the WZ prescriptions also works in the presence of the soft anomalies, we can ask if it is possible to interpret the resulting charges as improved Noether charges for some specific choice of boundary Lagrangian. 
The reason why this is not obvious is that since $\bar a_\xi\neq 0$, we have
\be\label{iNflux2}
\pbi{I_\xi\bar\th} = d \bar q_\xi + \bar a_\xi, 
\ee
where
\be
\bar q_\xi:=q_\xi+i_\xi b, \qquad \bar a_\xi := a_\xi+\D_\xi b.
\ee
This can be proved from \eqref{thbar}, or read off directly from \eqref{iNflux} using the fact that we can interpret $b$ as a boundary Lagrangian.\footnote{That fact that $b$ is only defined on the boundary, namely that $db\equiv 0$, does not affect the derivation.}
The compatibility of this equation with \eqref{WZflux} follows from \eqref{zidane}, and if  $\bar a_\xi\neq 0$ there is a mismatch.
 
Accordingly, we can distinguish three situations, depending on what anomalies are present:
\bit
\item[$(a)$] The preferred symplectic flux $\bar\th$ is associated to a total Lagrangian $L+d b$ without anomalies. Then $\bar a_\xi=0$, and we have
\be\label{WZa}
q^{\sscr WZ}_\xi =  \bar q_\xi = q_\xi + i_\xi b.
\ee 
In this case, the WZ charge coincides with an improved Noether charge with boundary Lagrangian $\ell = b$ and vanishing $\vth$.
The flux formula \eqref{WZflux} is consistent with \eqref{iNflux2} since the anomaly vanishes.

\emph{Remark:} covariance of both $\th$ and $\bar\th$ is not enough to guarantee $\bar a_\xi=0$, see \eqref{anomalies}.

\item[$(b)$] There are soft anomalies, and $C_\xi=0$. Then
\be
 \bar a_\xi =d s_\xi, \qquad \d s_\xi=-\bar q_{\d\xi}-A_\xi.
\ee
Then
\be\label{WZb}
q^{\sscr WZ}_\xi = q_\xi + i_\xi b + s_\xi.
\ee 
It differs from the improved Noether charge $\bar q_\xi$ that would be immediately associated with \eqref{thbar}, namely with boundary Lagrangian $\ell=b$ and $\vth=0$. 
Notice that the additional term $s_\xi$ is precisely the shift in the charge required so that the anomaly is removed from its flux, and this is how \eqref{iNflux2} is mapped to \eqref{WZflux}. When anomalies are present, the WZ prescription eliminates them from the flux, and puts them in the definition of the charge.

Next, we can ask if there exists a choice of boundary Lagrangian whose improved Noether charges reproduce the same shift. 
This is possible if $\bar q_\xi$ and $q^{\sscr WZ}_\xi$ are related by \eqref{iN}, namely if we can find a corner term that satisfies \eqref{sDc}, the same general equation of the previous Section applies here. If a solution to this equation exists, then the WZ
charge is the improved Noether charge $q^{\sscr c}$ with boundary Lagrangian $\ell^{\sscr c}=b+dc$. This fixes the corner ambiguity in the charges.
Notice that $\D_\xi\ell^{\sscr c}=-a_\xi$ and that $\ell^{\sscr c}$ is not unique, since any further shift by an anomaly-free corner term will also work, and produce the same charges.

\emph{Remark:} If we start from a covariant Lagrangian, $a_\xi=0$ and the required shift can be identified computing the anomaly of $b$,
\be
\bar a_\xi = \D_\xi b = ds_\xi.
\ee
As a consequence, the corner-improved boundary Lagrangian is covariant,
\be\label{totti}
\D_\xi \ell^{\sscr c}=\D_\xi b+d \D_\xi c = 0.
\ee
If furthermore the starting $\th$ is also covariant, then $A_\xi=0$ and
\be\label{interplay}
\d s_\xi=-\bar q_{\d\xi}.
\ee
This means that $\bar q_{\d\xi}$ is integrable, and exposes an interesting interplay that occurs between soft anomalies and field-dependent diffeomorphisms.
This interplay will be crucial below to understand why one can do calculations for the BMS group passing over anomalies.

From this analysis we deduce that when it is possible to reproduce the WZ charges as improved Noether charges, the latter can be identified a priori as those associated with a covariant choice of bulk \emph{and} boundary Lagrangians. All allowed anomalies can be restricted to corner terms.
Again $\ell^{\sscr c}$ is not unique, since adding anomaly-free corner terms will give the same charges. Therefore, it is enough to pick any representative in the class of anomaly-free Lagrangians.

\item[$(c)$] A general soft anomaly is present, including $C_\xi$. 
The flux of the improved Noether charge contains an extra term with respect to the WZ flux. The two equations are still compatible because \eqref{WZflux} is valid up to field-space constant terms. If $C_\xi$ is spacetime-exact it can be reabsorbed in the Hamiltonian charge, and the matching with an improved Noether charge can be obtained following the same analysis as case $(b$). If $C_\xi$ is not spacetime-exact, the matching is not possible. We could not find any examples in which this situation occurs, but we have no arguments to rule it out either. Lacking both, we refrain from drawing any conclusions about this case.

\eit

Notice that \eqref{interplay} can also be proven directly from Noether's theorem, as follows. For an arbitrary $\d\xi$ and assuming no anomalies in the initial Lagrangian and symplectic potential, we have
\begin{align}\label{interplay2}\nn
\pbi{dq_{\d\xi}}&=\pbi{I_{\d\xi} \th} -\pbi{i_{\d \xi} L} =I_{\d\xi} \pbi{\th} -\pbi{i_{\d \xi} L} 
=(\d_\xi - \mathcal{L}_\xi - \Delta_\xi) \pbi{\th} - \pbi{i_{\d \xi} L} \\
&= (\d_\xi - \mathcal{L}_\xi)(\bar \theta - \d b) - \pbi{i_{\d \xi} L} 
=-(\d_\xi-\pounds_\xi-\D_\xi)\d b - \pbi{i_{\d \xi} L}
=-\d\D_\xi b -\pounds_{\d\xi} b - \pbi{i_{\d \xi} L}.
\end{align}
If $\d\xi$ is now restricted to a symmetry vector, it is tangent and thus the last term vanishes. Using this and $\pounds_{\d\xi} b=d i_{\d\xi} b$, we recover \eqref{interplay}.
This alternative derivation highlights that a non-vanishing $q_{\d\xi}$ means that one is working with a symplectic potential that is not covariant, in spite of not being anomalous, because $I_{\d\xi}\th\neq 0$. 
In this case, the covariance requirement of the WZ flux means trading
\be
\D_\xi \th=0,\quad I_{\d\xi}\th\neq 0 \qquad \Rightarrow \qquad \D_\xi \bar\th=- I_{\d\xi}\bar\th\neq 0.
\ee

Summarizing, the WZ covariance requirement is enough to guarantee that the anomaly and field-dependent diffeomorphism contributions to \eqref{Ixiom} are integrable. The WZ prescription \eqref{WZdef} can be interpreted as an improved Noether charge constructed so to have the anomalous term $\bar a_\xi$ shifted from the flux to the definition of the charge. This shift can be identified a priori if it possible to find a covariant boundary Lagrangian. In other words, it is premature to conclude from \eqref{thbar} that $b$ is the boundary Lagrangian. If $\D_\xi b\neq 0$, one should rather look for a covariant $\ell^{\sscr c}=b+dc$.

This leads to the following independent definition of the WZ charges. First, evaluate \eqref{thbar}, choosing a covariant $\bar\th$, namely such that
\begin{subequations}\label{WZnew}\be
(\d_\xi - \mathcal{L}_\xi) \bar \theta =0,
\ee
plus the conservative or stationarity requirement chosen as described in Cases I and II.
Second, identify a corner term $c$ such that 
\be
\ell^{\sscr c}:=b+dc, \qquad \D_\xi\ell^{\sscr c}=0.
\ee\end{subequations}
Finally, compute the improved Noether charge associated with $(\bar\th,\ell^{\sscr c})$, or in other words with the split
\be\label{WZiN}
\pbi{\th}=\bar\th -\d\ell^{\sscr c}+d\d c.
\ee

\subsection{Extending the WZ prescription to non-field-exact corners}

This prescription can be immediately generalizing condition 0 to \eqref{th'}, as pointed out in \cite{Harlow:2019yfa}. All that will change is that the identification of $\ell^c$ will start from \eqref{th'} instead of from $b$. We repeat the procedure in this case, for the sake of clarity and ease of reference.

First, evaluate \eqref{th'}, choosing $\th'$ such that
\begin{subequations}\label{WZnew2}\be
(\d_\xi - \mathcal{L}_\xi)\th'=0,
\ee
plus the conservative or stationarity requirement chosen as described in Cases I and II. 
Second, identify a corner term $c$ such that 
\be
\ell^{\sscr c}:=\ell+dc, \qquad \D_\xi\ell^{\sscr c}=0.
\ee\end{subequations}
Finally, compute the improved Noether charge associated with $(\th',\ell^{\sscr c})$, or in other words with the split
\be
\pbi{\th}=\th' -\d\ell^{\sscr c}+d(\vth +\d c).
\ee
This means computing
\be\label{genWZc}
q^{\sscr c}_\xi= q'_\xi -\D_\xi c = q_\xi +i_\xi\ell^{\sscr c} -I_\xi \vth^{\sscr c},\qquad \vth^{\sscr c}:=\vth+\d c,
\ee
where 
$q_\xi$ is the Noether charge associated with the covariant bulk Lagrangian, and $q_\xi'$ is the Noether charge associated with the pair $(\th',\ell)$. The resulting flux is anomaly-free, $d q^{\sscr c}_\xi = I_\xi \th' -i_\xi L'$, as opposed to the anomalous flux of $q'_\xi$, given by \eqref{iNflux}.
Of course, if we choose directly an anomaly-free $\ell$, then $q_\xi'\equiv q_\xi^{\sscr c}$.

\section{Examples}

We review here some well-known explicit examples, which will be useful to provide some physical meaning to the anomaly $\bar a_\xi$.
The examples are all based on the Einstein-Hilbert Lagrangian, hence $a_\xi=0$. We always start from the bare potential, which is not anomalous,
\be
\th=\f1{3!} \th^\m \eps_{\m\n\r\s}dx^\n\w dx^\r \w dx^\s, \qquad \th^\m = 2 g^{\r[\s} \d \G^{\m]}_{\r\s} = 2g^{\m[\r}g^{\n]\s} \na_\n\d g_{\r\s}, \label{thg}
\qquad \D_\xi \th=0.
\ee
On the other hand, notice that $I_{\d\xi}\th$ needs not vanish for arbitrary $\d\xi$. Therefore, while this potential is covariant in the absence of field-dependent diffeomorphisms, in general we can only say that it is non-anomalous, and it may well be non-covariant. Indeed, we will see that in the case of asymptotic Killing vectors at $\scri$ with a field-dependent extension, this is not zero, and therefore the bare Einstein-Hilbert potential is not covariant.

For each example below, we review the WZ prescription, determine in which of the cases $(a,b,c)$ we are, and compute when needed the anomaly contribution to obtain the WZ charges as improved Noether charges. The first two examples concern case I, and the latter two concern case II.
As we will see the first three examples are somewhat trivial, therefore our discussion may appear slightly pedantic there. However we believe they allow us to explain our logic before the more involved fourth example, and also to provide a useful comparative of the literature.

\subsection{Conservative time-like boundary}\label{SecTimelike}

Recasting the pull-back of the bare Einstein-Hilbert symplectic potential in Dirichlet form on a time-like hypersurface $\cal T$ with normal $n_\m$, one finds 
(see e.g. \cite{Burnett:1990tww})
\be
\pbi{\th} 
= \Pi_{\m\n} \d q^{\m\n}  \eps_{\cal T} - \d\ell +d\vth,\qquad \Pi_{\m\n}:=K_{\m\n}-  q_{\m\n}K
\ee
where $\ell= 2K\eps_{\cal T}$ is the Gibbons-Hawking-York term. The explicit form of $\vth$ can be found in \cite{Burnett:1990tww,Iyer:1995kg} or more recent references, and will not be needed in the following. Suffices to say that it can be made to vanish with an appropriate choice of coordinates, corresponding to orthogonal corners \cite{Iyer:1995kg,Odak:2021axr}.\footnote{If the time-like boundary is at $r=$constant and the cross sections are defined by the space-like hypersurfaces with time-like unit normal $\t:= -N dt$, the restriction is a vanishing $r$-component of the shift vector. This means that the time-like boundary is orthogonal with all $t=$constant hypersurfaces, condition preserved by variations such that $\t_\m n_\n \d g^{\m\n} =0 $. This restriction implies $\vth=0$.}
Restricting to this situation, the non-integrable term in \eqref{thbar} is a good candidate for the preferred symplectic potential $\bar \th$. It satisfies condition 0 with $b=\ell$, and condition 2 with conservative boundary conditions $q^{\m\r}q^{\n\s}\d q_{\r\s}\stackrel{\cal T}=0$, or with a notion of stationarity given by $\Pi=0$.
To discuss its covariance, we evaluate
\be
(\d_\xi-\pounds_\xi)  \left(\Pi_{\m\n} \d q^{\m\n} \right) = \D_\xi  \Pi_{\m\n} \d q^{\m\n} + \Pi_{\m\n}\d \D_\xi q^{\m\n} 
+ \Pi_{\m\n} \pounds_{\d\xi} q^{\m\n}. 
\ee
The residual diffeomorphisms that preserve the phase space must preserve the boundary, hence be tangent to it. As a consequence 
$\d_\xi q_{\m\n}=\pounds_\xi q_{\m\n}$, and $\D_\xi n_\m = 0$ provided we work with a unit-norm normal (see Appendix~\ref{AppA}).
Therefore  
\be\label{timeanomaly}
\D_\xi q_{\m\n}=\D_\xi\eps_{\cal T}=\D_\xi K_{\m\n}=0.
\ee 
We conclude that this $\bar\th$ is covariant for field-independent diffeomorphisms.
The boundary symmetry group is Diff$(\cal T)$, and the charges will be conserved for arbitrary variations around solutions with $\Pi=0$,
and around arbitrary solutions but only for variations restricted  to preserve the boundary conditions. In the latter case the only allowed symmetries of the conservative subset of the phase space are the Killing vectors of the boundary metric.

Following the prescription used in \cite{Iyer:1995kg} and adopted in the WZ paper, we take conservative boundary conditions, as in Case I. 
Now that we have chosen $\bar\th$ and identified $b$ with the Gibbons-Hawking-York (GHY) term, the next step is to look at the anomalies. We have $a_\xi=0$ from the initial choice of the Einstein-Hilbert Lagrangian, and $\D_\xi b=0$ from \eqref{timeanomaly}. Therefore all anomalies vanish and we are in case $(a)$.
According to \eqref{WZa}, the WZ charge is given by the improved Noether charge $q_\xi+i_\xi b$. 
One can easily check that this is indeed the WZ charge computed in \cite{Iyer:1995kg,Wald:1999wa}, and which gives the Brown-York formulas at finite distance.

Let us also comment about the importance of the contribution of \cite{Harlow:2019yfa}.
If we relax the corner-orthogonality condition, we have $\vth\neq 0$ hence condition 0 is violated. This brings us outside of the hypothesis used in \cite{Iyer:1995kg,Wald:1999wa}. However conditions 1 and 2 are still valid.
The crucial insight of \cite{Harlow:2019yfa} was to show that the modification \eqref{omprime} of the symplectic two-form caused by $\vth$ is not only acceptable, but indeed leads to the correct Brown-York formulas in the case of non-orthogonal corners. 
This calculation is an example of the generalized WZ prescription \eqref{genWZc}, and the insight on the relevance of the redefined symplectic form $\om'$ played an important role in the general developments reviewed in Section~\ref{SecImproved}.

The WZ charges so obtained are still improved Noether charges with boundary Lagrangian $b$, namely case $(a)$, even with non-orthogonal corners. In fact, 
the presence of a second normal in $\vth$ will make some of the boundary diffeomorphisms anomalous, but \eqref{timeanomaly} still holds, and the GHY Lagrangian remains covariant even with non-orthogonal corners. Therefore no corner shift is needed to obtain the WZ charges. The BY formulas follow indeed from \eqref{WZa} with $b$ given by the GHY Lagrangian also with non-orthogonal corners \cite{Harlow:2019yfa,Odak:2021axr}.\footnote{This should be compared with the $3+1$ canonical calculation with conservative boundary conditions \cite{Brown:2000dz,Harlow:2019yfa,Odak:2021axr}, where the 2d Hayward corner term is needed in order to obtain the BY formulas with non-orthogonal corners. Its role is to secure the right Legendre transform on the boundary.
There appears to be no relation between the Hayward term in the action and the corner shift \eqref{qc} in the covariant improved Noether charge, which we use only to remove non-covariance from the boundary Lagrangian and satisfy the WZ conditions, and is not needed here. 
}

Finally, one can consider different $\bar\th$'s associated with other boundary conditions, and this leads to a modification of the Brown-York formulas \cite{Odak:2021axr}.

\subsection{Spatial infinity}

Lack of radiation makes the case of spatial infinity trivial. 
Using ADM fall-off conditions, we have 
\begin{equation}\label{i0}
\pbi\th=\d b, \qquad b=\lim_{r\rightarrow\infty}\left(\left(\partial_r g_{t t}-\partial_t g_{r t}\right)+r^c q^{ab}\left(\partial_a q_{bc}-\partial_c q_{ab}\right)\right)\epsilon_{\cal T},
\end{equation}
and $\eps_{\cal T}$ is the volume element on the time-like slices of constant $r$. The residual diffeomorphisms with non-trivial charges correspond to the asymptotic Poincar\'e group, and regarding them as limits of the analysis at finite distances shows that there are no anomalies. Hence we can take $\bar\th\equiv 0$.
Spatial infinity corresponds to case I in the WZ paper, namely the pull-back of the symplectic 2-form on the lateral boundary vanishes identically. 
From \eqref{i0} we see that the WZ prescription trivializes, since the Hamiltonian generator is manifestly integrable, with charge given by
 $q_\xi+i_\xi b$ \cite{Iyer:1994ys}, and as  shown there it reproduces the ADM formulas.

From the improved Noether charge perspective, this is just like the case at finite distances, we are in case $(a)$ since $\D_\xi b=0$, hence the formula coincides with 
taking $b$ as boundary Lagrangian. While it is not immediately clear which boundary conditions are identified by this choice, the limiting procedure obtained starting from a time-like boundary at finite distance shows that $b$ corresponds to Dirichlet boundary conditions up to a renormalization term, see e.g.  \cite{Odak:2021axr}.
The renormalization term depends on a chosen Minkowski background and therefore can potentially introduce anomalies, however these vanish because of \eqref{timeanomaly} and the restriction to Poincar\'e transformations of the asymptotic symmetries.

It would be interesting to see if the situation changes relaxing boundary conditions as to have non-trivial super-translations and super-rotation charges, as investigated in \cite{Henneaux:2018hdj}.
Finally, we mention that leaky boundary conditions at infinity (with non-vanishing cosmological constant) have been studied in \cite{Compere:2020lrt}.

\subsection{Finite null hypersurfaces and NEH}
Restricting the variations to preserve the universal structure defined in  \cite{Chandrasekaran:2018aop}, one has
\be
\pbi{\th} = \big(\s^{\m\n} - \f\th2 
\g^{\m\n}\big)\d\g_{\m\n}\eps_\cN +2\d(\th\eps_\cN),
\ee
where $\s^{\m\n}$ and $\th$ on the RHS are the shear and expansion of the null hypersurface, and we hope that no confusion arises from the use of the same letter as for the bare symplectic potential.
The non-integrable first term is the preferred $\bar\th$ put forward in  \cite{Chandrasekaran:2018aop}, and satisfies all WZ requirements: covariance restricts the variations in phase space to preserve the normal and its inaffinity (see \cite{Chandrasekaran:2018aop} and the related discussion in \cite{nBYus}), and field-independence of the symmetry vector fields on the boundary;\footnote{It requires $\d k=0$ because otherwise $\bar\th$ depends on $k$ which in turns depends on the representative chosen and not on the universal structure. This in fact is because it is not class III invariant.} the stationarity requirement is satisfied for arbitrary variations around hypersurfaces with vanishing shear and expansion, which are related to non-expanding horizons (NEH) \cite{Ashtekar:2021kqj}. 
It follows that 
\be
b = - 2 \th\eps_\cN = -2d\eps_S. 
\ee
One can check that this quantity has vanishing anomaly \cite{Chandrasekaran:2020wwn,nBYus}, namely $\D_\xi b=0$. Furthermore, $a_\xi=0$ since we started from the covariant Einstein-Hilbert Lagrangian, hence $\bar a_\xi=0$. We conclude that we are in case $(a)$, the WZ charge coincides with the improved Noether charge with $\ell=b$ and $\vth=0$.

Notice that even if the boundary Lagrangian is covariant, anomalies are present, and it has been shown that this charge is different that what one could call the null Brown-York tensor \cite{Chandrasekaran:2021hxc}. Finally, alternative choices of stationarity and their effect on charges are explored in \cite{nBYus}.

\subsection{Future null infinity}

At future null infinity, using for instance Bondi coordinates,\footnote{For descriptions with geometric quantities only and avoiding reference to Bondi coordinates, see e.g. \cite{Ashtekar:1981bq,Ashtekar:2014zsa,Grant:2021sxk}.}
one has 
\be\label{thScri}
\pbi{\th} = -
\Big(2\d M
-\f12 \d(D_AD_BC^{AB})
+\f12N_{AB}\d C^{AB} -\f18\d(N_{AB}C^{AB})\Big)\eps_\scri.
\ee
See Appendix~\ref{SecBMS} for definitions and some details.
At a first sight, one may identify the physical flux with the non-integrable third term, namely the Bondi news $N_{AB}:=\p_u C_{AB}$ contracted with the variation of the shear $C_{AB}$.
The stationarity requirement is then satisfied by all spacetimes with vanishing news, for arbitrary variations.
The issue though is that this term does \emph{not} satisfy the covariance requirement, because $N_{AB}$ is not covariant. 
The resolution of this issue was found by Geroch \cite{Geroch:1977jn} with the introduction of a background tensor $\Ger_{AB}$ carrying his name, and whose transformation property is $\D_\xi \Ger_{\la AB\ra}=\D_\xi N_{AB}$, so that
\be
\hat N_{AB}:=N_{AB}-\Ger_{\la AB\ra}
\ee
is covariant. 
The Wald-Zoupas criteria thus single out as preferred potential  \cite{Wald:1999wa}\footnote{In \cite{Wald:1999wa}, the covariant news $\hat N$ are denoted $N$, referred to as Bondi news, and one keeps in mind that the expression $\p_u C_{AB}$ is only valid in the special set of Bondi frames. This is indeed a better nomenclature in our opinion. We maintain however the $N$ and $\hat N$ notation here to match more easily with the contemporary literature, where Geroch's analysis seems to have been forgotten at some point. Notice also that $\hat N$ can be defined in geometric terms as the Lie derivative of the shear \cite{Grant:2021sxk}, which makes its covariance manifest. Geroch's construction on the other hand used the Schouten tensor of the (pull-back of the) unphysical Riemann tensor, hence a non-covariant quantity, and which coincides with $\p_u C_{AB}$ in Bondi coordinates.}
\be\label{barthBMS}
\bar\th = -
\f12\hat N_{AB}\d C^{AB}\, \eps_\scri.
\ee
The remainder is a total variation and identifies
\be\label{bscri}
b= 
\Big(2 M
-\f12\bar D_A\bar D_B C^{AB}-\f18 N_{AB}C^{AB}   +\f12\Ger_{AB} C^{AB}   \Big)\eps_\scri.
\ee
Therefore all three conditions for the WZ prescription are met.

A key property of $\Ger_{\la AB\ra}$ is to vanish identically when the background metric is the round 2-sphere.
This makes the choice $N\d C$ numerically correct in such Bondi frames, however one should keep in mind that the correct potential secretly depends on the Geroch tensor in order to secure covariance. This is relevant for us, because calculation of the anomaly involves derivatives in field space, and the anomaly of $b$ would be different if we forgot the term necessary to the covariance of $\bar\th$.

From \eqref{bscri} and the anomaly-freeness of the Einstein-Hilbert Lagrangian we compute 
\be\label{sxiscri}
\bar a_\xi = \D_\xi b = d s_\xi, \qquad {\rm where} \qquad s_\xi :=\f14 C^{AB} {\bar D}_A{\bar D}_B\t,
\ee
and $\t:=\xi^u=T+\tfrac u2 D_AY^A$. Details of this calculation are in the Appendix.
We see that we are in case $(b)$: there is a shift, caused by the fact that the `naive' boundary Lagrangian $b$ has an anomaly. 

The shift can furthermore be obtained from the corner ambiguity. In fact, using the anomalous transformations reported in Appendix~\ref{SecBMS}, it is easy to find a local functional $c$ solving \eqref{sDc}:
\be\label{BMScorner}
c:=\f1{16}C_{AB}C^{AB}\eps_S, \qquad \D_\xi c = -s_\xi.
\ee
Therefore the BMS charges obtained from the WZ prescription can also be obtained as improved Noether charges,
choosing an anomaly-free boundary Lagrangian such as
\be\label{ellscri}
\ell^{\sscr c}=b+dc = \Big(2 M
-\f12\bar D_A\bar D_B C^{AB}  +\f12\Ger_{AB} C^{AB}   \Big)\eps_\scri, \qquad \D_\xi\ell^{\sscr c}=0.
\ee
This choice of boundary Lagrangian is of course not unique: any further shift by an anomaly-free  corner term would give the same charges. 
In this case, it means that we can add an arbitrary contribution proportional to 
\be\label{trivialc}
(N_{AB}C^{AB}-2{\bar D}_A{\bar D}_B C^{AB})\eps_\scri, 
\ee which is both a corner term and anomaly-free.
In other words, one can equivalently use any element in the family
\be\label{ellx}
\ell^{\sscr c}_x = \Big(2 M
-\f{1+x}2\bar D_A\bar D_B C^{AB} +\f x4N_{AB}C^{AB} +\f12\Ger_{AB} C^{AB}   \Big)\eps_\scri, \qquad x\in\R.
\ee

Summarizing, the WZ charges for the BMS group can be obtained as improved Noether charges following the prescription \eqref{WZnew}, namely starting from the covariant Einstein-Hilbert Lagrangian and bare symplectic potential, and choosing $(\bar\th,\ell^{\sscr c})$ given by \eqref{barthBMS} and \eqref{ellx} respectively.
The resulting corner term in \eqref{WZiN} is \eqref{BMScorner}.

The fact that the the BMS charges can be obtained as improved Noether charges is consistent with what observed in \cite{Freidel:2021yqe,Chandrasekaran:2021vyu}, where relevant boundary Lagrangians were identified a posteriori.
The novelty of our derivation is the identification of the boundary Lagrangians and charges from first principles, thanks to the attention paid to anomalies.\footnote{
When comparing our quantitative results with the literature, some attention is however needed. The corner term \eqref{sxiscri} differs from the one used in \cite{Freidel:2021yqe} by a factor of 2 (mind the different units used, $16\pi G=1$ here, $8\pi G=1$ there). This follows from the fact that \cite{Freidel:2021yqe} uses the tetrad Lagrangian, whose bare symplectic potential differs from the Einstein-Hilbert one by a corner term \cite{DePaoli:2018erh,Oliveri:2019gvm}. We explain this comparison in App.~\ref{AppTetrads}, and our results here perfectly agree with those of \cite{Freidel:2021yqe}. We find on the contrary a disagreement with the conclusions of \cite{Chandrasekaran:2021vyu}, which find (1) no anomalous shift, namely they claim $q^{\sscr WZ}_\xi=\bar q_\xi+i_\xi b$, and (2) no restriction of the boundary Lagrangian to be anomaly-free, namely they consider four different options of which only their (6.17) is in our family, with $x=0$. The remaining (6.9), (6.18) and (6.18) are anomalous. 
Disagreement (1) is in our opinion due to a computational mistake, we believe that their equation (6.15) lacks a factor 1/2 in the third term, the one that reads ${\cal U}{\cal D}f$. Their numerical factor would indeed make the anomalous shift unnecessary in order to obtain the BMS charges, but it is in contradiction with our calculation reported in App.~\ref{SecBMS}, as well as with the calculations of \cite{Barnich:2011mi,Compere:2018ylh}
 which use the $q_{\d\xi}$ term, and which we report in App.~\ref{AppArcheo}. Since the presence of a non-zero anomaly is crucial to our paper, we made multiple checks of our calculations and the presence of this contribution. But of course we welcome further feedback on this point, should the mistake be on our end.
Disagreement (2) is on the other hand not an issue, provided (1) is fixed. Every time their boundary Lagrangian is anomalous, they redefine the charge by hand to remove what the anomalous contribution to the flux, via the quantity they denote $\tl h_\xi$. What we have shown here is that $\tl h_\xi$ is in general not an improved Noether charge in the sense of \eqref{iN}, and that there is no need to do this redefinition by hand, because it is possible to identify the charge uniquely working with a covariant pair of bulk and boundary Lagrangians.
}

As a final remark, the WZ-compatible anomaly \eqref{sxiscri} has the structure of the soft term  in the flux-balance laws for the BMS charges.
This example provides a physical example of the meaning of the anomaly contribution to the variation of the improved Noether charge \eqref{iNflux}: 
an improved Noether charge with boundary Lagrangian \eqref{bscri} as opposed to \eqref{ellscri}, or with an arbitrary corner improvement not selected by the covariance requirement \eqref{totti}, would differ from the standard BMS charges by soft terms. One consequence would be that they would measure different memory effects, another that the boost part of the charges would fail to be conserved on stationary spacetimes.
The relation between the soft terms and anomalies is further explained by the detailed calculations reported in the Appendix \ref{SecBMS}, which highlight how the bridge between the two lies in the first-order extension of the symmetry vector fields. We also report there the calculation of the charges (App.~\ref{SecBMS}), as well as the explanation of how \eqref{interplay} allows one to do the calculation \`a la Wald-Zoupas without the need to take explicitly into account  the anomalies (App.~\ref{AppArcheo}).

\section{Conclusions}

We have studied the WZ prescription in the light of the recent extensions of the covariant phase space.
The WZ prescription introduced two precise and valuable physical requirements, spelled by conditions 1 and 2 in Section~\ref{SecWZ}. First, the importance of covariance with respect to the background structure, and second, the importance of a physical notion of stationarity in the radiative case. 
On the other hand, the WZ prescription appears to be too restrictive concerning condition 0, which forbids allowing changes in the corner potential, the quantity we referred to as $\vth$ in this paper. The importance of changing the corner potential was stressed in \cite{Harlow:2019yfa} and elaborated further in \cite{Freidel:2020xyx} and subsequent literature. In particular, it is necessary in case I if one wants to recover the Brown-York charges with non-orthogonal corners at finite distance\cite{Harlow:2019yfa,Odak:2021axr}, and in case II if one wants to consider generalizations of the BMS group \cite{Campiglia:2020qvc}.
Our discussion hopefully highlights the importance of taking physical principles, as opposed to a mathematical prescription alone, in selecting the symplectic potential defining the charges. This is possibly the key lesson of the WZ paper. 

The first result that emerges from our study is that the WZ prescription works also in the presence of anomalies and field-dependent diffeomorphisms. These cannot be arbitrary, but are restricted from the covariance requirement \eqref{WZfootnote} to satisfy the condition \eqref{zidane}. We referred to the allowed anomalies as soft because of their physical meaning at future null infinity. This example also shows that the bare Einstein-Hilbert symplectic potential is not covariant, in spite of  being non-anomalous, because it is sensitive to the possibly field-dependent extensions of the asymptotic Killing vectors. The WZ requirement of covariance balances this dependence with an anomaly as in \eqref{WZfootnote}.

The second result is that the WZ charges are not straightforwardly improved Noether charges in the sense of \cite{Harlow:2019yfa}, namely they cannot necessarily be obtained from the formula \eqref{iN}. It only happens if \eqref{sDc} admits solutions. This is the case for the BMS charges, but we don't know if it is always possible.
It would surely be interesting to further study this differential equation and understand its general solution. There may also be hybrid situations in which the WZ charges are improved Noether charges only for a subset of the boundary symmetry algebra.

When  \eqref{sDc} can be solved, the WZ charges are improved Noether charges, up at most to field-constant terms.
Solving that equation has the compelling meaning that one has to find a boundary Lagrangian that is anomaly-free, when starting from a covariant bulk Lagrangian. In other words, the boundary Lagrangian needed to derived the WZ charges from \eqref{iN} is not necessarily the term $b$ that appears in \eqref{thbar}, nor the initial $\ell$ that appears in \eqref{th'}. 

This understanding allows us to provide an independent definition of WZ charges as the improved Noether charges satisfying \eqref{WZnew}, and to generalize it relaxing condition 0 by requiring \eqref{WZnew2}.

A non-trivial property of the soft anomalies is to be strongly related to the allowed field-dependent diffeomorphisms, via 
\eqref{zidane} in general, and via \eqref{interplay} if one starts from a covariant bulk Lagrangian and its bare potential. A consequence of this is that one can effectively perform some calculations ignoring anomalies, and this is the reason why Wald and Zoupas could compute the BMS charges without the need to talk about anomalies. Nonetheless, we believe it does not affect the relevance of taking anomalies into account. We hope that our new derivation of the BMS charges in the main text and in the Appendix shows that using the formalism with anomalies enriches our understanding of the mathematics as well as the physics. 

One example of what can be learnt is how anomalies capture the difference between future null infinity and a null hypersurface at a finite distance, such as an event horizon. It is well known that the BMS symmetries are different from the symmetries of a null hypersurface in spacetime. For example, in BMS, dilations are not independent while on a finite null hypersurface they are independent and their associated charge is given by the area.\footnote{Matching the two symmetries is possible relaxing the fall-off conditions so that the BMS group is enhanced to the BMSW group \cite{Freidel:2021yqe}. See also \cite{Donnay:2016ejv,Donnay:2019jiz,Adami:2021nnf} for related work on charges at horizons.}
In both cases the anomaly comes from the boundary normal. At a finite distance, the background structure only provides the location of the boundary. But at future null infinity, it also provides the compactification factor. As a consequence, the metric on the cross-section is anomaly-free at finite distance, but not on $\scri$. This introduces a second source of anomalous transformations, given by the inhomogeneous terms of the metric functionals on $\scri$.
It is also interesting to remark that the difference can be seen looking at the first-order extension of the symmetry vector fields. More details on these aspects appear in Appendix~A and D.

\subsection*{Acknowledgments}
We thank Anthony Speranza for discussions on anomalies, Adrien Fiorucci, Roberto Oliveri and Ali Seraj for discussions on BMS charges,
and Luca Ciambelli for comments on the draft.

\appendix

\section{Anomalies and boundaries}\label{AppA}
In this Appendix we review how to compute the anomaly associated with a background structure, and prove the absence of anomalies in the case of time-like boundaries parametrized by a unit-normal, the result used in \eqref{timeanomaly}.

Anomalies arise when the covariant phase space contains background structures. 
Let us denote by $\phi$ the dynamical fields, and by $\chi$ the background fields.
For the dynamical fields we define $\d_\xi \phi:=\pounds_\xi \phi$, whereas the background fields satisfy $\d_\xi \chi=0$, whence the anomaly $\D_\xi \chi = -\pounds_\xi\chi$.
To understand the third term in \eqref{Dxidef}, consider a functional of the fields that is a one-form in field space, namely $F(\phi,\chi)\d\phi$. In this case we have
\be
\d_\xi (F\d\phi) 
= \p_\phi F\d_\xi \phi \d\phi +F\d\d_\xi\phi = \p_\phi F\pounds_\xi \phi \d\phi +F\d\pounds_\xi\phi
= \pounds_\xi (F\d\phi) -\p_\chi F\pounds_\xi\chi \d\phi + F\pounds_{\d\xi}\phi,
\ee
where we used $[\d,\d_\xi]=0$ in the first equality, and $[\d,\pounds_\xi]=\pounds_{\d\xi}$ in the last. Hence,
\be
\D_\xi (F\d\phi) = -\p_\chi F\pounds_\xi\chi \d\phi = (\d_\xi-\pounds_\xi-I_{\d\xi})F\d\phi.
\ee

The first example of background structure we consider in the following is a spacetime boundary $\cal B$. We define it by its Cartesian equation as $\chi(x^\m)=0$, and associated with it a normal 1-form $n_\m:=-f\p_\m\chi$. The field $\chi$ is a fixed background structure, with $\d\chi=0$. Since $\pounds_\xi\chi=\xi^\m\p_\m\chi$, every diffeomorphism that does not preserve the boundary is anomalous. When constructing the covariant phase space associated to this boundary, the only relevant diffeomorphisms are those that preserve the boundary, namely
\be\label{bdiff}
\xi^\m n_\m\stackrel{\cal B}=0 \qquad \Rightarrow \qquad \xi^\m=\bar\xi^\m+\chi\hat\xi^\m,
\ee
where $\bar\xi^\m n_\m=0$. The boundary is shared by all metrics in the phase space. The diffeomorphisms that preserve the boundary are also called residual diffeomorphisms, or symmetry vector fields, hinting at the physical relevance that boundary diffeomorphisms can acquire.
In different situations, one may add additional background structure on top of the presence of the boundary, still shared by all metrics in the phase space and usually referred to as \emph{universal} structure. Any additional requirement in the universal structure can restrict the symmetry group.

From \eqref{bdiff} it follows that $\pounds_\xi\chi\stackrel{\cal B}=0$, and therefore $\D_\xi \chi=0$: the boundary is covariant with respect to the diffeomorphisms that preserve it. However, anomalies can still appear when we look at  derivatives of $\chi$, for instance through the normal 1-form. In fact, a simple calculation shows that
\be\label{Dn}
\D_\xi n_\m = w_\xi n_\m, \qquad w_\xi:=\D_\xi \ln f-\hat\xi^\m\p_\m\chi.
\ee
If we take a gradient as normal, say $f=1$, then the anomaly comes entirely from $\hat\xi^\m\p_\m\chi$, namely from how much the extension of $\xi$ off $\cal B$ does not preserve the neighbouring leaves of the $\chi$ foliation. 
However, as long as the foliation is not null, the anomaly associated with a non-trivial extension $\hat \xi$ can be eliminated choosing $f$ so that the normal is unit-norm: in this case in fact,
\be
n_\m = s\f{\p_\m\chi}{\sqrt{sg^{\r\s}\p_\r\chi\p_\s\chi}}, \qquad n^2=s:=\pm1,
\ee
and
\begin{align}\nn
\D_\xi n_\m &=- s\f{1}{\sqrt{sg^{\r\s}\p_\r\chi\p_\s\chi}}\left(\pounds_\xi\p_\m\chi -sg^{\n\l}\f{\p_\n\chi \pounds_\xi\p_\l\chi }{sg^{\r\s}\p_\r\chi\p_\s\chi}\p_\m\chi\right) \\
&=- s\f{1}{\sqrt{sg^{\r\s}\p_\r\chi\p_\s\chi}}\left(\d^\n_\m-s n^\n n_\m\right)\pounds_\xi\p_\n\chi =-q^\n_\m\pounds_\xi n_\n = 0
\end{align}
because of the condition that $\xi$ preserves the boundary.
Recalling that unit-norm means foliation independence of the normal, we see that what this anomaly is capturing is not so much the presence of the boundary, but rather any foliation-dependence in its description, namely non-invariance under $\chi\mapsto\chi'(\chi,x^\m)$.

In the case of a null hypersurface, there is no choice of $f$ that would make the normal foliation-independent, hence anomalies \eqref{Dn} are generically present.
Furthermore, in order to distinguish physical solutions on the covariant phase space, one typically reduces the allowed variations to preserve a certain universal structure \cite{Wald:1999wa,Chandrasekaran:2018aop}. This reduces the symmetry group and can lead to a fixed, non-vanishing first-order extension, hence anomalies. 
 An interesting difference arises between a null hypersurface at a finite distance and future null infinity. In both cases, we have a background field describing the presence of the boundary. But at future null infinity, the same structure is used as conformal factor $\Om$ in the compactification. As a consequence, reparametrizing $\chi$ at finite distance changes the normal 1-form $n_\m$, but reparametrizing $\Om$ changes both the normal and the unphysical metric which induces the metric on the cross-sections, leading to two sources of anomalies. To see this difference in formulas, consider the volume elements. At finite distance we have
\be
\eps_{\cal N}=i_l \eps= -l\w \eps_S,
\ee
where $l\cdot n=-1$ is the auxiliary vector,
hence \eqref{Dn} implies \cite{Chandrasekaran:2020wwn}
\be
\D_\xi\eps_{\cal N} = -w_\xi \eps_{\cal N}, \qquad \D_\xi \eps_S =0.
\ee
But the volume element of future null infinity is determined from the unphysical metric $\tl g_{\m\n}:=\Om^2 g_{\m\n}$, which is anomalous:
\be
\D_\xi \Om = 0, \qquad \D_\xi n_\m = w_\xi n_\m, \qquad \D_\xi \tl g_{\m\n} = 2w_\xi \tl g_{\m\n}, \qquad \D_\xi\tl\eps=4w_\xi\tl\eps.
\ee
Therefore taking 
\be
\eps_{\scri}=i_l \tl\eps= -l\w \eps_S,
\ee
we have
\be\label{Dxiepsscri}
\D_\xi\eps_{\scri} = 3w_\xi \eps_{\cal N}, \qquad \D_\xi \eps_S =2w_\xi\eps_S.
\ee
The first difference is that the anomalous dimension of the induce volume element changes from $-1$ to $+3$, and the second difference is that tensors on the cross-sections are now anomalous as well, unlike in the finite dimensional case.
This comes as explained above from the fact that the background structure has the double role of determining the boundary and providing the unphysical metric.

\section{BMS anomalies}\label{SecBMS}

We review here some basic formulas of the BMS transformations, and show how to compute the associated anomalies and the shift between the WZ and improved Noether charges. We follow \cite{Freidel:2021yqe} for the notation. While the general logic remains the same described in the main text, performing the calculations explicitly requires paying attention to two special features. The limit to $\scri$ and the difference between a symmetry vector field on $\scri$ and its bulk extension, and the fact that we choose to work with  a specific coordinate system.
Working in Bondi coordinates $(u,r,\th,\phi)$ and with conformal factor $\Om:=1/r$, 
the asymptotic Killing vectors are 
\be\label{xiBMS}
\xi := \t\p_u +Y^A\p_A+ \Om(\dot\t\p_\Om - \p^A\t\p_A) +O(\Om^2).
\ee
For the BMS group, $\t=T+\tfrac u2 D_AY^A$, where $T(\th,\phi)$ is the supertranslation parameter, and $Y^A(\th\,\phi)$ a conformal Killing vector on the two-sphere. For the BMSW enlargement \cite{Freidel:2021yqe}, which encompasses both extended \cite{Barnich:2010eb} and generalized \cite{Campiglia:2014yka,Compere:2018ylh} BMS groups, $\t=T(\th,\phi)+u W(\th,\phi)$, and $Y^A$ is an arbitrary vector, which we take to be globally defined. 
From 
\be
\D_\xi \tl g_{\m\n} = -g_{\m\n}\pounds_\xi \Om^2 = -\f2{\Om}\xi^\Om \tl g_{\m\n} = -2\dot\t \tl g_{\m\n},
\ee
we see that 
\be\label{wxiscri}
w_\xi = -\dot\t.
\ee

The covariant phase space at future null infinity is parametrized by the functionals
$\bar q_{AB}, C_{AB}$, respectively the leading and first sub-leading orders of the 2d metric, and the mass and angular momentum aspects $M, \bar P_A$. The parametrization is chosen so that $\bar P_A$ coincide with the definition of Dray and Streubel in these coordinates. All quantities depend on $(u,\th,\phi)$, except for the background metric $\bar q_{AB}$ which is constant in $u$.\footnote{With Penrose's definition of asymptotic flatness (see e.g. \cite{Wald:1999wa,Ashtekar:2014zsa}), one can always choose a conformal factor satisfying the Bondi condition $\tl\na_\m n^\m=0$, and then $\pounds_n \bar q_{AB}=0$. This is the case with the choice of $\Om$ taken here, from which the asymptotic Einstein's equations impose $\p_u\bar q_{AB}=0$.}
The phase space transformations generated by the asymptotic BMSW symmetries are \cite{Freidel:2021yqe}
\begin{subequations}\begin{align}
& \d_{\xi} \,\bar{q}_{A B}= (\pounds_Y-2 \dot{\tau}) \bar{q}_{A B}, \\
& \d_{\xi} \,C_{A B}= (\t\p_u + \pounds_Y -\dot{\tau}) C_{A B}-2 \bar{D}_{\langle A} \partial_{B\rangle} \tau,  \\
& \d_{\xi} \,N_{A B}= (\t\p_u + \pounds_Y)N_{AB} -2 \bar{D}_{\langle A} \partial_{B\rangle} \dot{\tau},  \\
& \d_{\xi} \,M= (\t\p_u + \pounds_Y+3 \dot{\tau}) M+\left(\f{1}{2} \bar{D}_A N^{A B}+\p^B \bar{F}\right) \p_B \tau +\f{1}{4} \p_u(C^{A B} \bar{D}_A \p_B \tau), \\
& \d_{\xi}\, \bar{P}_A= (\t\p_u + \pounds_Y+2 \dot{\tau}) \bar{P}_A 
+3 M \partial_A \tau-\frac{1}{8} N_{B C} C^{B C} \partial_A \tau+\frac{1}{2}\left(C_A^C N_{B C}\right) \partial^B \tau \\\nn
&\qquad+\frac{3}{4}\left(\bar{D}_A \bar{D}_C C_B{ }^C-\bar{D}_B \bar{D}_C C_A C\right) \p^B \tau+\frac{1}{4} \p_A\left(C^{B C} \bar{D}_B \bar{D}_C \tau\right) \\\nn
&\qquad+\frac{1}{2} \bar{D}_{\langle A} \bar{D}_{B\rangle} \tau \bar{D}_C C^{B C}+C_{A B}\left(\bar{F} \partial^B \tau+\frac{1}{4} \partial^B \Delta \tau\right).
\end{align}\end{subequations}
Here $\pounds_Y$ is a slight abuse of notation and should be understood as the Lie derivative for tensors on the two-sphere. 
The functionals transform in general not as scalars but rather as densities in the $u$ variable, because of the $\dot\t$ shifts, and as tensors on the sphere in the $A$ indices, plus inhomogeneous terms. 
Because of this algebraic structure, $\scri$ is endowed with the structure of a fiber bundle $S^2\times \R$ in which the fibers are the conformal weights. 
The density shifts and the inhomogeneous terms are responsible for the anomalies.

To see that explicitly, we need first to explain how the covariant Lie derivative is mapped to a gauge-fixed description associated with the Bondi coordinates used above. Consider a 3-form on $\scri$. This is a spacetime covariant quantity, which in Bondi coordinates will read like a scalar on the 2-sphere times the volume form, e.g. $v_A w^A \eps_\scri$. For an example, see the symplectic potential at $\scri$ given by \eqref{thScri}. Using the asymptotic symmetry vectors \eqref{xiBMS}, we have
\be
\pounds_\xi (v_A w^A \eps_\scri) = \pounds_\xi (v_A w^A)\eps_\scri + v_A w^A\pounds_\xi \eps_\scri.
\ee
Now we can write
\be\nn
\pounds_\xi (v_A w^A)=\xi^\m \p_\m (v_A w^A) = \t\p_u (v_A w^A)+Y^B\p_B(v_A w^A)=
\t\p_u (v_A w^A)+\pounds_Y(v_A w^A)= L_\xi (v_A w^A),
\ee
where we introduced the Bondi-frame Lie derivative
\be
L_\xi:=\t\p_u +\pounds_Y, 
\ee
or Bondi Lie derivative for short. Its action is that of a Lie derivative on the conformal bundle of 
$u$-dependent 2-sphere tensors.

The anomalies of the phase space functionals are thus given by $\D_{\xi}=\d_{\xi}-L_{\xi}$
(the last term from the definition \eqref{Dxidef} drops out because we are acting on field-space scalars), 
 namely 
\begin{subequations}\label{BMSanomalies}\begin{align}\label{Dxiq}
& \D_{\xi} \,\bar{q}_{A B}= -2 \dot{\tau} \bar{q}_{A B}, \\
& \D_{\xi} \,C_{A B}= -\dot{\tau} C_{A B}-2 \bar{D}_{\langle A} \partial_{B\rangle} \tau,  \\
& \D_{\xi} \,N_{A B}= -2 \bar{D}_{\langle A} \partial_{B\rangle} \dot{\tau}, \label{DxiN} \\
& \D_{\xi} \,M= 3 \dot{\tau} M+\left(\f{1}{2} \bar{D}_A N^{A B}+\p^B \bar{F}\right) \p_B \tau +\f{1}{4} \p_u(C^{A B} \bar{D}_A \p_B \tau),
\end{align}\end{subequations}
and similarly for $\bar P_A$, which won't be needed in the following. These formulas are identical for BMSW and BMS, with the only simplification for BMS being that $\dot\t=\bar D_AY^A/2$, and that taking the round sphere $\bar F=\bar R/4=1/2$, so one term in $\D_\xi M$ drops out.
From now on, we restrict attention to the BMS case.

The anomaly of the background metric is familiar from the BMS literature: the 2d metric -- aka `Bondi frame' -- is a background structure, hence $\d \bar q_{AB}=0$, while an asymptotic symmetry changes the Bondi frame by a conformal transformation given by $2\dot\t=\bar D_A Y^A$.
Hence the RHS of \eqref{Dxiq}.
In the generalized BMS and in BMSW the phase space is enlarged to include the Bondi frame as a variable, hence $\d\bar q_{AB}\neq 0$, but the resulting anomaly is again just a conformal transformation, albeit with an arbitrary factor instead of just the Lorentz boost $\bar D_AY^A$.
From this expression we can also derive $\D_\xi\sqrt{\bar q} = \sqrt{\bar q} \bar q^{AB}\D_\xi\bar q_{AB}/2=-2\dot\t\sqrt{\bar q}$.

The anomaly of the news \eqref{DxiN} is also familiar from the seminal work of Geroch \cite{Geroch:1977jn}, which introduced the tensor carrying his name, and whose traceless part is
\be
\Ger_{\la AB\ra}:=2\bar D_{\la A} \varphi \bar D_{B\ra} \varphi + 2\bar D_{\la A} \bar D_{B\ra}\varphi,  
\ee
where $2\varphi$ is the conformal factor relating the metric $\bar q_{AB}$ to a round 2-sphere metric. It vanishes for the round 2-sphere. From this expression and the condition $\p_u\bar q_{AB}=0$ we deduce that $\pounds_\xi \r_{\la AB\ra}=2\bar D_{\la A} \bar D_{B\ra}\dot\t$. Notice that it is crucial that $\r_{AB}$ has a trace part that does not vanish on a round 2-sphere, otherwise this Lie derivative would vanish as well. Geroch also proves that this tensor is universal. Hence $\d\Ger_{\la AB\ra}$=0, and
\be
\D_{\xi} \,\Ger_{A B}= -2 \bar{D}_{\langle A} \partial_{B\rangle} \dot{\tau} \equiv \D_\xi N_{AB}.
\ee
It follows that
$\hat N_{AB} := N_{AB}-\Ger_{AB}$
is covariant, i.e. its anomaly vanishes.

The anomaly of the volume form $\eps_\scri$ is given by
\eqref{Dxiepsscri} with \eqref{wxiscri}, namely\footnote{It is also possible to derive this writing
$\eps_\scri =du\w \eps_S$. The 1-form $du$ is an anomalous quantity on the scale bundle, with anomaly given by 
$\D_\xi du=-\pounds_\xi du = -\dot\t du$, and $\D_\xi \eps_S =-2\dot\t\eps_S$. Care is needed when writing $\eps_S=\sqrt{\bar q}d^2\th$ and using the anomaly for $\sqrt{\bar q}$ previously derived. This is because $\sqrt{\bar q}$ is a density, therefore we should remember that $d^2\th$ is an invariant. This is a familiar result for Lie derivatives of volume forms: if we write $\eps=\sqrt{-g}d^4x$, where $d^4x:=\f1{n!}\ut \eps_{\m\n\r\s}dx^\m\w d^\n\w dx^\r\w dx^\s$ is a density, we have 
$\pounds_\xi\sqrt{-g}=\sqrt{-g}\na_\m\xi^\m$ and $\pounds_\xi d^4x=0$.}
\be\label{Dxieps}
\D_\xi \eps_\scri = -3\dot\t\eps_\scri.
\ee
Putting together these results, we find
\begin{align}\nn
(\d_\xi - \pounds_\xi) \left(\hat N_{AB} \d C^{AB} \eps_{\scri} \right) 
&= (\d_\xi - L_\xi) \hat N_{AB} \d C^{AB} \eps_{\scri} + \hat N_{AB} (\d_\xi - L_\xi)\d C^{AB} \eps_\scri + \hat N_{AB} \d C^{AB} (\d_\xi - L_\xi)\eps_\scri \\\nn
&= \D_\xi \hat N_{AB} \d C^{AB} \eps_{\scri} + \hat N_{AB} \d\D_\xi C^{AB} \eps_\scri + \hat N_{AB} L_{\d \xi} C^{AB} + \hat N_{AB} \d C^{AB} \D_\xi \eps_\scri \\
&= \hat N_{AB} \d \D_\xi  C^{AB} \eps_\scri -  3\dot\t \hat N_{AB} \d C^{AB} \eps_\scri =0.
\end{align}
In the third equality we used $L_{\d \xi} C^{AB} = 0$, since $\d\xi=O(\Om^2)$ as follows from \eqref{xiBMS}.
In the last we used $\D_{\xi} \,C^{A B}= 3\dot{\tau} C^{A B}-2 \bar{D}^{\langle A} \partial^{B\rangle} \tau$ which follows from \eqref{BMSanomalies}. 
As a consequence, the non-integrable term that can be read naively from \eqref{thScri} is not covariant, whereas $\bar\th$ defined in \eqref{barthBMS} is.

Next, we compute the anomaly of \eqref{bscri}, here copied for convenience:
\be\label{bApp}
b= 
\Big(2 M+D_A\bar U^A -\f18 N_{AB}C^{AB}   +\f12\Ger_{AB} C^{AB}   \Big)\eps_\scri,
\ee
with $\bar U^A= -\f12\bar D_B C^{AB}$.
For this term, we have\footnote{The second equality below follows from the fact that for BMS
\be\nn
[\D_\xi, \bar D_A] = -[\pounds_\xi, \bar D_A],
\ee
which can be computed from
\begin{align}\nn
& [\t\p_u,\bar D_A]f^B = -\p_u f^B \bar D_A\t, \\ \nn
& [\cL_Y,\bar D_A]f^B= f^A \bar D_B \bar D_CY^C -\f R2(2f^B Y_A-\d^B_A Y_C f^C), \qquad [\cL_Y,\bar D_A]f^A= f^A \bar D_A \bar D_CY^C ,
\end{align}
and we also observe that $[\p_u,\cL_Y]=0.$}
\begin{align}\nn
\D_\xi (\bar D_A\bar U^A \eps_\scri) &= \bar D_A(\D_\xi \bar U^A) \eps_\scri + \bar D_A\bar U^A\D_\xi \eps_\scri +[\D_\xi,\bar D_A]\bar U^A \eps_\scri \\
&\nn = \bar D_A(\D_\xi \bar U^A) \eps_\scri -3\dot\t \bar D_A\bar U^A \eps_\scri +(\dot{\bar U}^A\p_A\t -2\bar U^A\p_A\dot\t) \eps_\scri \\
 & = \bar D_A\big((\D_\xi -3\dot\t)\bar U^A\big) \eps_\scri +\f12 \p_u(C^{AB}\bar D_A\bar D_B\t).\label{DUan}
\end{align}
The total derivatives on the sphere can be dropped. As a consequence, we don't need to know the explicit form of the anomaly of $\bar U^A$. For the interested reader, it can be found in \cite{Freidel:2021yqe}.
For the other terms in $b$, we have
\be
\D_\xi \Big((2 M-\f18 N_{AB}C^{AB}   +\f12\Ger_{AB} C^{AB}  )\eps_\scri \Big) =
 -\f14 \p_u(C^{AB}\bar D_A\bar D_B\t) +\f12\Ger_{AB} \D_\xi ( C^{AB} \eps_\scri) .
\ee
Adding up and using the vanishing of  the Geroch tensor  on the round 2-sphere, we conclude that
\be\label{sApp}
\D_\xi b = ds_\xi, \qquad s_\xi = \f14 C^{AB}\bar D_A\bar D_B \t   \, \eps_S.
\ee
This proves \eqref{sxiscri} used in the main text.\footnote{Notice that the contribution of the density weights to the anomaly drops out, and it is only the inhomogeneous terms of \eqref{BMSanomalies} that matter in the end. Hence the calculations are consistent with those of \cite{Freidel:2021yqe}, where the densities were not included in the definition of the anomaly.}
As for \eqref{BMScorner}, this follows immediately computing the anomaly of $\bar\b:=-\tfrac1{32}C_{AB}C^{AB}$, which gives 
\be\label{cscri}
c:=-2\bar\b\eps_S, \qquad \D_\xi c = -\f14 C^{AB}  \bar D_A\bar D_B\t \,\eps_S.
\ee

Let's check that the shift \eqref{sxiscri} indeed reproduces the known expressions of the WZ charges at $\scri$. The expansion of the Komar two-form gives
(to lighten the notation, we drop in the following the sphere indices $A$)
\be
q_\xi = \t(2M- D\bar U+\f1{8}CN-\f14DDC)+2Y(-r\bar U+\bar P+\p\bar\b)+2W(-r^2+2\bar\b).
\ee
The divergent terms vanish for BMS, and can be renormalized away for BMSW \cite{Freidel:2021fxf}, so we will drop them in the following.\footnote{Taking the on-shell value of $\bar U$, the second and fourth terms add up to $-\tfrac12\bar D\bar U$. Our $\bar U$ coincides with the $\cal U$ used in \cite{Chandrasekaran:2021vyu}, and our 1/2 instead of their 1 in the third term of their (6.15) is the mismatch we referred to in the main text.}
Then we have
\be
i_\xi\eps_\scri=-\xi\cdot l \, \eps_S, \qquad  -\xi\cdot l = \t, \qquad 
i_\xi b = \t\Big(2 M+\bar D\bar U -\f18 NC   +\f12\Ger C   \Big)\eps_S,
\ee
hence
\be
q_\xi+i_\xi b = \left(\t(4M -\f14\bar D\bar DC)+2Y\bar P\right)\eps_S. 
\ee
This shows that without the right corner shift, the improved Noether charge with $b$ as boundary Lagrangian doesn't give the standard BMS charges. The difference is a soft term. This charge would not measure the standard memory effects, and furthermore has a flux determined not only by the physical symplectic potential $\bar\th$, but by the anomalous contribution as well. In particular, the part of the charges corresponding to Lorentz boosts would not be conserved in stationary spacetimes.

Finally, adding up \eqref{sApp}, which after a trivial integration by parts on the 2-sphere can be rewritten as
\be\label{sDU}
s_\xi = -\f\t2\bar D\bar U\eps_S,
\ee
we obtain the desired result\footnote{Recall we are using the notation from \cite{Freidel:2021yqe} and units $16\pi G=1$. The relation to the angular momentum aspect used in \cite{Barnich:2011mi,Compere:2018ylh} is $N_A=\bar P_A+\p_A\bar\b$.}
\be\label{BMScharges}
q^{\sscr WZ}_\xi=q_\xi+i_\xi b +s_\xi = \left(4\t M +2Y\bar P\right)\eps_S.
\ee
These charges vanish exactly on the Minkowski solution, therefore there is no need of any shift by field-space integration constants.
The calculation proves that the WZ charges can be obtained without ever talking about Hamiltonian generators, but just as an improved Noether charge with the prescriptions \eqref{WZnew}. The anomaly-free boundary Lagrangian can be read from \eqref{cscri} to be \eqref{ellscri}, which we report here for convenience,
\be\label{ellscriApp}
\ell^{\sscr c} = \left(2M+\bar D\bar U+\f12\Ger C\right)\eps_\scri.
\ee
We also notice that
\be\label{trivialcApp}
\left(\bar D\bar U+\f14 CN\right)\eps_\scri = d\left(\f18 C^2\eps_S + i_{\bar U}\eps_\scri 
\right),\qquad \bar U^\m := (0,0, \bar U^A).
\ee 
This corner term is also anomaly-free once we integrate on the 2-sphere to get rid of the total derivatives that appear when using \eqref{DUan}. 
We conclude that the WZ charges can be obtained starting from the family of boundary Lagrangians \eqref{ellx}, that all differ from \eqref{ellscriApp} by a term proportional to \eqref{trivialcApp}.

\section{Charges' archeology}\label{AppArcheo}

In this Appendix we comment on the importance of the interplay relation \eqref{interplay}. This allows one to understand how Wald and Zoupas were able to get away without ever talking about anomalies, and will also be the opportunity for us to add some comments about \cite{Barnich:2011mi,Flanagan:2015pxa,Grant:2021sxk} that we think may be useful to the reader. If we start from the Einstein-Hilbert Lagrangian there are no anomalies, and
\be\label{IxiomA}
-I_\xi\om=\d q_{\xi} -q_{\d\xi}-i_\xi\th.
\ee
Then, the WZ prescription \eqref{WZdef} gives
\begin{align} \label{WZ3}
-I_\xi\om+i_\xi\bar\th&=\d q_{\xi} -q_{\d\xi}-i_\xi\th+i_\xi\bar\th =\d q_{\xi} -q_{\d\xi}+i_\xi \d b \\
&=\d(q_\xi+i_\xi b)-q_{\d\xi} - i_{\d\xi}b.\nn
\end{align}
If we take the bare Eistein-Hilbert $\th$, this is covariant and $q_\xi$ is Komar; the covariance requirement for $\bar\th$ guarantees not only \eqref{zidane} but also \eqref{interplay}. Therefore,
 $\d s_\xi = -\bar q_{\d\xi} =-q_{\d\xi} - i_{\d\xi}b$. Using this equality in \eqref{WZ3} we recover the calculation of the charges done at the end of the previous Section, namely adding $s_\xi$ as computed from the anomaly of $b$.
But we can also forget about the anomalous origin of $s_\xi$, and compute directly $q_{\d\xi}$ and $i_{\d\xi} b$ in \eqref{WZ3}. 
On first thought, one may imagine that these vanish, since there is no field dependence in $\xi$ at zeroth or first order, see \eqref{xiBMS}. However, it had been observed as early as \cite{Geroch:1981ut} that the limit of the Komar 2-form to future null infinity depends on the second-order extension as well, and in fact it even depends on the \emph{third} order insofar as the radial component is concerned. This can be trivially checked using for instance Bondi coordinates and $\Om=1/r$. The Komar formula then contains $\p_r\xi^r$, which when integrated against the $r^2$ area 2-form fishes a contribution $O(r^{-1})$ in $\xi^r$, which is $O(\Om^3)$. But then, the second and higher-order terms are generically field-dependent. 
Using the Tamburino-Winicour extension, equivalent to preserving the bulk Bondi coordinates used in the previous Section, we have
\be
\d\xi = \left( \f{\Om^2}2\d(C^{AB}\p_B\t) +O(\Om^3)\right) \p_A  +\left(\f{\Om^3}2\d({\bar D}_A C^{AB}\p_B\t+\f12C^{AB}\bar D_A\p_B\t)+ O(\Om^4)\right)\p_r.
\ee
This vector gives a vanishing contribution when hooked with $b$, but not when plugged in the Komar form. There, it replaces a divergent term that was a total divergence on the sphere (hence integrating to zero) if $\xi$ was used, with a finite term that is no longer a total 2d-divergence, but rather gives on the cross-sections
\be\label{qdxiscri}
-q_{\d\xi} = \f14 \d(\t\bar D\bar D C )\eps_S.
\ee
This is precisely the same contribution of $s_\xi$, as expected from the general equivalence \eqref{interplay}. 
As a consequence, one can do the calculation using the first line of \eqref{WZ3}, and obtain the correct result without ever talking about anomalies, and instead properly taking into account the $q_{\d\xi}$ term \eqref{qdxiscri}. 
This is the way the calculation is done for instance in \cite{Compere:2018ylh}, even though the contribution of the term \eqref{qdxiscri} is not explicitly reported.\footnote{We thank Adrien Fiorucci for sharing his calculations.}
Notice also that the neat result of this term is to make $I_\xi\om$ independent of the field-dependent extension, because \eqref{qdxiscri} cancels the $O(\Om^3)$ term that appears when computing $\d q_\xi$, As for the second-order terms in $q_\xi$, they drop out when taking the pull-back on a fixed $u$ cross-section of $\scri$.
The final result depends only on the zeroth and first orders of $\xi$, which are field-independent. 

This term is also taken into account in the formula used in \cite{Barnich:2011mi}, following \cite{Barnich:2001jy,Barnich:2004uw}, and this is for us the only reference in the literature where all aspects of the calculation of the BMS charges are properly and explicitly discussed.\footnote{Mind however that \cite{Barnich:2011mi} does not start from $-I_\xi\om$ but adds to it a term proportional to the Killing equation, see e.g. (9.10) in \cite{Oliveri:2019gvm}. This additional term has vanishing limit to $\scri$.}

Coming back to the WZ paper, there are actually two difficulties with the way the BMS calculations are presented. The first is that since they assume $\d\xi=0$, they write $-I_\xi\om=\d q_{\xi} -i_\xi\th.$ This is not too bad, because it can be easily corrected:
the effective consequence of the $q_{\d\xi}$ term in \eqref{WZ3} is that one should take the variation of $q_\xi$ treating $\xi$ as a $c$-number even if it is field-dependent. With this caveat in mind, the calculations are correct. Otherwise, (94) of \cite{Wald:1999wa} is missing an additional finite term coming from the $O(\Om^2)$ terms of $\xi$.
Notice that WZ discuss the independence of \eqref{IxiomA} from the arbitrary part of the extension of the asymptotic symmetry vector, below their equation (22). This independence is taken there as a definition of equivalent representatives, but it can be proved explicitly as done in \cite{Grant:2021sxk}, Lemma 5.2. The proof is given there only for field-independent higher-order extensions, but can be trivially generalized to our case if $\d q_\xi$ is replaced by $\d q_\xi-q_{\d\xi}$. Or alternatively, with the caveat that $\xi$ is always a $c$-number for $\d$.
This way of understanding the action of $\d$ and the RHS of $I_\xi\om$ for field-dependent diffeomorphisms was made more explicit shortly after in \cite{Gao:2003ys}. 
We suppose that this is the approach taken also in \cite{Grant:2021sxk}, even though it is nowhere explicitly stated, otherwise some of their calculations are missing intermediate terms that cancel out in the end result. 
We remark that having extended the proof of independence from higher-order extensions to the field-dependent case, one can also compute the RHS of \eqref{IxiomA} 
ignoring such terms, instead of computing them and see that they cancel out. This means in particular ignoring the $q_{\d\xi}$ term altogether. This provides another way of interpreting the results of of \cite{Wald:1999wa,Grant:2021sxk} as correct.
With these caveats in mind, \cite{Grant:2021sxk} is a very clear and explicit paper, and has the further advantage of presenting the calculations in two different gauges as well as in covariant language. 

The second difficulty of the WZ paper concerns the boost charges. Inspection of \eqref{xiBMS} shows that these get a contribution from the vertical part, and therefore are not generated purely by a vector tangential to the cross-section. In other words, restricting $\xi$ to be tangential is a stronger condition than setting the super-translation parameter to zero. Nonetheless, Wald and Zoupas tried to recover all Lorentz charges, rotations as well as boosts, from a purely tangential vector. The interest in doing so is possibly that for a field-independent and purely tangential vector, the Hamiltonian generator is integrable since the pull-back of $i_\xi\th$ vanishes, and one does not need any prescription. The result is the Komar formula, which they knew gives the Dray-Streubel charges for angular momentum but not for boosts, unless the extension is chosen to satisfy the Geroch-Winicour condition.  So what Wald and Zoupas set up to do is to prove that the variation of the Komar formula is unchanged if the Geroch-Winicour condition is imposed, because then they can claim that the Dray-Streubel charges are recovered when they further impose the condition that all charges vanish in Minkowski spacetime.
This is arguably a more tortuous path than straightforwardly including the vertical part of the vector in the boost contribution, which is the reason the calculation works no matter what extension is taken.

The same result of \cite{Barnich:2011mi} then appeared again in \cite{Flanagan:2015pxa}. Both papers use the Tamburino-Winicour extension described above. However \cite{Flanagan:2015pxa} claims that the result matches Dray-Streubel because the Geroch-Winicour condition $\na_\m \xi^\m=0$ can be relaxed to $\na_\m \xi^\m=O(\Om^2)$, which is satisfied by the Tamburino-Winicour extension that they use. This argument is wrong in our opinion, because the Tamburino-Winicour extension precisely requires the linkage term in order to reproduce the right boost charges \cite{Tamburino:1966zz}. The reason why \cite{Flanagan:2015pxa} gets the right charges is for us not that the linkage is not needed because of the chosen extension, but because of the correct inclusion of the vertical term, just as in \cite{Barnich:2011mi}.

As a final comment, notice that \eqref{qdxiscri} shows that $I_{\d\xi}\th\neq 0$ for the Einstein-Hilbert bare potential, by consistency with the Noether theorem \eqref{iNflux} with $\xi$ replaced by $\d\xi$. 
To verify this explicitly some care is needed, because $\d\xi$ is \emph{not} a symmetry vector.\footnote{It vanishes on $\scri$, and the Tambourino-Winicour extension of the trivial vector on $\scri$ vanishes everywhere, unlike $\d\xi$. The latter is more akin to the difference between two different bulk representatives of the same asymptotic Killing vector.}
In particular, $\d_{\d\xi}$ does not exist on the asymptotic phase space. 
 Instead, we can use the general formula \eqref{interplay2} and take the limit to infinity. Since $\d\xi=O(\Om^2)$, the last two terms vanish and we find
\be
\lim_{r\rightarrow\infty}\pbi{I_{\d\xi} \th} = -\d\D_\xi b = -d\d s_\xi.
\ee
This result together with $q_{\d\xi}=-\d s_\xi$ proves the consistency of \eqref{qdxiscri} with Noether's theorem.

\section{Anomalies and first-order extensions of symmetry vector fields}

In Appendix A we showed that the case of future null infinity differs from a finite distance null hypersurface because there are two sources of anomalies.
We point out that this difference is encoded also at the level of the asymptotic Killing vectors, if one looks at the first-order extension away from the boundary. 
For a null-hypersurface at finite distance, located say at $r=0$,  we have \cite{Chandrasekaran:2018aop} 
\be
\xi = \t\p_u +Y^A\p_A - r\dot\t \p_r  +O(r^2).
\ee
At future null infinity we have \eqref{xiBMS}, which we report here for convenience of comparison:
\begin{align}
\xi& = \t\p_u +Y^A\p_A + \Om(\dot\t\p_\Om - \p^A\t\p_A) +O(\Om^2)
\end{align}
In both cases the first-order is fixed uniquely in terms of the symmetry parameters, and the freedom to extend the symmetry vector field starts at second order. We see that the first-order extensions contain respectively one and two terms, and these are the seeds of the anomalous transformations: $\dot\t$ at a finite distance, whereas on $\scri$ we have both the density-weights $\dot\t$ as well as the inhomogeneous transformations that go like $\p_A\t$.

Another difference concerns the fact that field-dependent extensions of the diffeomorphisms don't matter in computing the charges at finite distance, but matter at $\scri$.
This is because the Komar formula depends on first derivatives of $\xi$, which are field-independent at finite distance, but involve higher orders at $\scri$, which see the field-dependence.

\section{Tetrad variables}\label{AppTetrads}

There are three useful remarks to make if one uses tetrad variables. First, the bare symplectic potential differs from the Einstein-Hilbert one by an exact 3-form \cite{DePaoli:2018erh,Oliveri:2019gvm}. Second, if one fixes the same physical $\th$ and boundary Lagrangian, the improved Noether charge is the same \cite{G:2021xvv}. Furthermore, the DPS exact 3-form is anomaly-free, therefore one can use the same covariant boundary Lagrangian as in the metric case to evaluate the Wald-Zoupas prescription for the BMS charges.  

The bare symplectic potential given by the Einstein-Hilbert Lagrangian differs from the tetrad one by an exact 3-form \cite{DePaoli:2018erh,Oliveri:2019gvm},
\be
\th = \th^e+d\a^{\sscr DPS}, \qquad \a^{\sscr DPS} = \star (e_I\w\d e^I).
\ee
As a consequence, the bare Noether charges computed without adding any boundary Lagrangian are also different, and we have
\be
q_\xi = q_\xi^e +I_\xi\a^{\sscr DPS},
\ee 
where $q_\xi$ is Komar, and $q_\xi^e=\tfrac12\eps_{IJKL}e^I\w e^J \,i_\xi\om^{KL}$.
The improved Noether charges can be made to coincide if one chooses the boundary Lagrangian $\ell$ and $\th'$ to match the metric choices, as pointed out in \cite{G:2021xvv}:
\be
q^{e\prime}_\xi = q^{e}_\xi +i_\xi\ell-I_\xi \vth^e = q_\xi +i_\xi\ell-I_\xi \vth = q'_\xi.
\ee
This is a perfect example of the value of working with the improved Noether charge, ambiguities such as picking a representative of the equivalence class become irrelevant once attention is switched to the physically preferred symplectic potential.\footnote{When the authors of \cite{Freidel:2020xyx} write the table of different corner symmetry algebras associated with the ADM, EH, EC and ECH Lagrangians, they are looking at the bare Noether charges $q_\xi$ associated with the bare symplectic potential and no boundary Lagrangian, as selected by the homotopy prescription. Should they switch to the improved Noether charges $q'_\xi$ selected in each case by the same $\th'$ and the same $\ell$, they would of course obtain the same algebra in each case.}

Both $\th^e$ and $\a^{\sscr DPS}$ are anomaly-free. Furthermore, $\a^{\sscr DPS}$ becomes field-space-exact at $\scri$, 
\be\label{dascri}
\pbi{d\a}^{\sscr DPS} = \d(DU+\f18CN).
\ee
Therefore condition 0 of the WZ prescription is satisfied, with 
\be
\pbi{\th^e} = -\Big(\d(2M+2D\bar U) + \f12 N_{AB}\d C^{AB} \Big)\eps_\scri.
\ee
Taking the same $\bar\th$ as before, we have
\be
b^e = \Big(2 M+2 D\bar U +\f12\Ger C  \Big)\eps_\scri.
\ee
The anomaly of this boundary Lagrangian can be computed as shown before and gives twice the metric one, $\D_\xi b^e=d s^e_\xi = 2d s_\xi$ with $s_\xi$ given by \eqref{sDU}.
On the one hand, this is the right result to get the correct WZ charge, since using the results of \cite{Freidel:2021yqe},
\be
q_\xi^{\sscr e}+i_\xi b^e = \left(\t(4M+D\bar U)+2Y\bar P\right)\eps_S, \qquad q_\xi^{\sscr e}+i_\xi b^e+2s_\xi = \left(4\t M+2Y\bar P\right)\eps_S.
\ee
On the other hand, this means that the corner shift needed to get this result from an improved Noether charge is also twice the metric one, 
\begin{align}
& c^e = 2c= -4\bar\b\eps_S= \f18 C^2\eps_S, \\
& \ell^e = b^e+d c^e = \Big(2 M+2 D\bar U +\f14 CN +\f12\Ger C  \Big)\eps_\scri, \qquad \D_\xi \ell^e=0.
\end{align}
Notice that this anomaly-free boundary Lagrangian differs from the metric one \eqref{ellscri} by the anomaly-free corner term  \eqref{trivialcApp}.
It thus belong to the same anomaly-free class, and can indeed be recognized as \eqref{ellx} with $x=1$.
That it belongs to the same family of anomaly-free boundary Lagrangians was to be expected, since $\D_\xi \a^{\sscr DPS}=0$.

In the same anomaly-free class of tetrad boundary Lagrangians we find, taking $x=-1$,
\be
\ell^{\sscr BMSW} = \Big(2 M - \f14 CN +\f12\Ger C  \Big)\eps_\scri,
\ee
which is the one used in \cite{Freidel:2021yqe}. Those results are thus perfectly compatible  with the ones here presented, and the novelty is that we now know how to identify this boundary Lagrangian a priori, without having to deduce it from already knowing the WZ charges. 

As a final remark, notice that there is no incompatibility between the fact that  \eqref{trivialcApp} and \eqref{dascri} are both anomaly-free in spite of having different relative factors, because $[\D_\xi,\d]=-\D_{\d\xi}\neq 0$. This calculation cannot however be done explicitly without providing a definition for $\d_{\d\xi}$, which in turns requires an extension of $b^e$.

\providecommand{\href}[2]{#2}\begingroup\raggedright\endgroup

%

\end{document}